\documentclass[pop,
 amsmath,amssymb,
 reprint,%
author-year,%
]{revtex4-1}

\usepackage{graphicx}
\usepackage{dcolumn}
\usepackage{bm}

\usepackage[utf8]{inputenc}
\usepackage[T1]{fontenc}
\usepackage{mathptmx}
\usepackage{etoolbox}

\usepackage[capitalize]{cleveref}
\usepackage{xcolor}
\usepackage{comment}
\includecomment{notestocoauthors}
\excludecomment{notestocoauthors}

\usepackage{soul}

\newcommand{\new}[1]{{#1}}
\newcommand{\old}[1]{}

\makeatletter
\def\@email#1#2{%
 \endgroup
 \patchcmd{\titleblock@produce}
  {\frontmatter@RRAPformat}
  {\frontmatter@RRAPformat{\produce@RRAP{*#1\href{mailto:#2}{#2}}}\frontmatter@RRAPformat}
  {}{}
}%
\makeatother
\begin{document}

\preprint{AIP/123-QED}

\title[Comparing $L$-dependent Radial Diffusion Ensembles]{Two methods to analyse radial diffusion ensembles: the peril of space- and time- dependent diffusion}
\author{S. N. Bentley}
\affiliation{Department of Maths, Physics and Electrical Engineering, Northumbria University.}%
\affiliation{Department of Meteorology, University of Reading.}
 \email{sarah.n.bentley@northumbria.ac.uk}

\author{J. Stout}
\affiliation{Department of Meteorology, University of Reading.}

\author{D. J. Ratliff}
\affiliation{Department of Maths, Physics and Electrical Engineering, Northumbria University.}%

\author{R. Thompson}
\affiliation{Department of Meteorology, University of Reading.}

\author{C. E. J. Watt}
\affiliation{Department of Maths, Physics and Electrical Engineering, Northumbria University.}%

\date{\today}

\begin{abstract}

Particle dynamics in Earth's outer radiation belt can be modelled using a diffusion framework, where large-scale electron movements are captured by a diffusion equation across a single adiabatic invariant, $L^{*}$ $``(L)"$. While ensemble models are promoted to represent physical uncertainty, as yet there is no validated method to analyse radiation belt ensembles. Comparisons are complicated by the domain dependent diffusion, since diffusion coefficient $D_{LL}$ is dependent on $L$. We derive two tools to analyse ensemble members: time to monotonicity $t_m$ and mass/energy moment quantities $\mathcal{N}, \mathcal{E}$. We find that the Jacobian ($1/L^2$) is necessary for radiation belt error metrics. Components of $\partial\mathcal{E}/\partial t$ are explicitly calculated to compare the effects of outer and inner boundary conditions, and loss, on the ongoing diffusion. Using $t_m$, $\mathcal{N}$ and $\mathcal{E}$, we find that: (a) different physically motivated choices of outer boundary condition and location result in different final states and different rates of evolution; (b) the gradients of the particle distribution affect evolution more significantly than $D_{LL}$; (c) the enhancement location, and the amount of initial background particles, are both significant factors determining system evolution; (d) loss from pitch-angle scattering is generally dominant; it mitigates but does not remove the influence of both initial conditions and outer boundary settings, which are due to the $L$-dependence of $D_{LL}$. We anticipate this study will promote renewed focus on the distribution gradients, on the location and nature of the outer boundary in radiation belt modelling, and provide a foundation for systematic ensemble modelling.


\end{abstract}

\maketitle

\begin{notestocoauthors}
    
    {\color{purple}
\section*{Plain Language Summary}
Earth's radiation belts are a region of highly charged particles, trapped by Earth's magnetic field. These particles pose a hazard to spacecraft and so there is considerable interest in forecasting this environment. On longer timescales, these particle motions can be simplified to a simple diffusion equation towards and away from the Earth (radial diffusion). However, to date there has been no clear method to compare simulations and therefore to deal with the uncertainty and variance that is both an unavoidable aspect of the underlying physics, and a hindrance to space weather forecasting. We explore two ways to compare these simulations and identify which aspects of the initial model, when changed, result in the system evolving differently. The two methods used are time to monotonicity and mass- and energy -like quantities, which we use to investigate the evolution of a background radiation belt plus an enhancement from locally energised particles, and to estimate a radial diffusion ``timescale''. We find that different choices of outer boundary condition and location result in a different final state and a different rate of evolution. We find that the gradients of the particle distribution affect the evolution more significantly than the spatial dependence of the diffusion coefficient. We find that the location of the enhancement, and the amount of background particles existing before the enhancement, are both significant factors determining how the system evolves. We discuss what this means for radiation belt modelling going forward, and make recommendations.

\section*{Message to editor}
atypical, authentic structure (not shoehorning to fit narrative). Also this is good practice when writing any other document? So we should do it here too.

Reflecting cycic practice of research

Ask their opinion on first paragraph to aid navigation, could move? could make an initial zero section, "Navigation"?

}
\end{notestocoauthors}

\section{Introduction}\label{sec:Intro}
\emph{Radial diffusion is a phenomenon studied in both space and fusion plasmas. In this work, we investigate how the radial dependence of that diffusion interacts with initial and boundary conditions. We provide here a guide for different readers to navigate this paper. Firstly, in the introduction we provide a broad introduction of the motivations behind understanding the uncertainty in radial diffusion modelling in near-Earth space. Plasma physicists without a radiation belt background may wish to review the application of radial diffusion in near-Earth space in \cref{sec:Background}. For all readers, we explicitly list our goals in \cref{sec:goals}, introducing labels which are used throughout the manuscript as we develop each research question, find relevant results and then discuss our conclusions. 
\cref{sec:Methods} contains details of the numerical models and develops the properties we require in an analysis metric, thereby motivating our use of time to monotonicity and the  mass- and energy-like quantities $\mathcal{N},\mathcal{E}$. We find and present the most significant ways in which these quantities vary in \cref{sec:results}. Radiation belt modellers and space weather physicists may be particularly interested in the suggestions we make for future based on our findings in \cref{sec:Discussion}, where we also compare our results to current modelling practices. Some of our more significant conclusions relate to the importance of the outer boundary and the particle gradients instead, which are also discussed in \cref{sec:Discussion}. Where possible, each section is self-contained, to enable those with different interests to find the relevant sections useful.}






Earth's radiation belts are a region of highly energised particles, magnetically trapped by the Earth's magnetosphere. The trapped particles undergo several types of periodic motion, the slowest of these being the drift around the Earth. Electromagnetic perturbations on timescales of the drift of electrons around the Earth will scatter those electrons onto orbits closer to, or more distant from, the Earth; this is radial diffusion. \new{Drift timescales can vary with particle energy from minutes to days but is typically considered to be on the order of a few hours.} Radial diffusion is one of the major drivers of Earth's radiation belts and due to the variation in timescales of radiation belt particles motions, an approximation for radiation belt modelling can be made using solely this mechanism \new{to reproduce the broad dynamics, although not shorter timescale phenomena such as dropouts \cite{Tu2014,Ozeke2017}}. This simple yet effective model enables us to explore ways of including uncertainty in our radiation belt models using ensembles. A full model of the radiation belts would require us to acknowledge that they are part of a complex, interdependent system, with consequently larger amounts of uncertainty. Radial diffusion modelling using the Fokker-Planck formalism is reviewed in more detail in \cref{sec:Background}; briefly, radiation belt modelling \new{using Fokker-Planck simulations} does not use the motion of individual particles but instead averages over motion on larger scales, where wave-particle interactions across many scales are included using diffusion coefficients. Where the wave-particle interactions are well understood, variability should be properly characterised in order to capture the correct uncertainty. Where these interactions are not well understood, \new{large uncertainties}\old{uncertainty across orders of magnitude} can indicate inadequate modelling. \new{We see uncertainty across orders of magnitude when estimating $D_{LL}$ in units of $days^{-1}$. For example, using different $D_{LL}$ models on diffusion estimated by particle tracing in an MHD simulation spans four orders of magnitude at $L=5$ (i.e. spanning minutes to days)\cite{Huang2010b}, while empirical models parameterised by geomagnetic activty still vary by more than one order of magnitude even at the same location and geomagnetic activity level \cite{Liu2016}. Unsurprisingly, varying the spatiotemporal distribution of diffusion coefficients changes the final particle distribution \cite{Thompson2020}. Previous work to address the uncertainty in diffusion coefficients shows that probabilistic inference of these coefficients outperforms our current empirical models \cite{Sarma2020}. Uncertainty can be both inherent to the system and an indicator of a poorly described system; both these situations can be addressed using appropriate modelling techniques.}  \old{For example, estimates of $D_{LL}$ under similar conditions vary by orders of magnitude [Huang2010b,Liu2016,Thompson2020,Sarma2020]} 


Ensembles can be used to quantify uncertainty in radiation belt modelling. Ensemble modelling involves running a simulation multiple times, with a variety of input conditions or parameter settings, to represent the unknowns in a given system. The impact of these unknowns on the final state can be quantified; alternatively, variability across model outputs provide a measure of uncertainty on that final state. Probability distribution of model outputs provide us with a better understanding of the uncertainty associated with our models. 
%
%
There are several sources of uncertainty in radiation belt modeling, including uncertainty due to approximations or physical unknowns, uncertainty in observations (i.e. in the measured value and in the spacecraft location), uncertainty in boundary conditions or model settings and physical uncertainty inherent to the system. Physical uncertainty arises from the fact that deterministic models may exhibit chaotic behaviour if they are particularly sensitive to initial conditions. \old{Given that the magnetosphere is a complex system and that we have very sparse observations}\new{Given that the magnetosphere is a complex system and that observations are spare compared to spatiotemporal scales of interest}, it is extremely likely that we will need some way to include this chaotic deterministic character. Furthermore, the computational requirements of our modelling mean that we have sub-grid physics: physics on smaller scales that must be included, but cannot be fully modelled numerically. Uncertainty in driving parameters, such as \old{upscale}\new{upstream} solar wind and ultra-low-frequency \new{(ULF)} waves driving radial diffusion, can also affect the accuracy of the model. \new{Example sources of uncertainty include the properties of the solar wind and how this drives ULF waves. In turn, the uncertainty in magnitude and spatiotemporal occurrence of ULF waves will result in uncertainty in radiation belt models. In addition to this uncertainty chain, modelling of each physical process adds further sources of uncertainty, for example to include the impact of ULF waves, it is typically assumed that the ULF azimuthal mode number is $m=1$, which does not hold in in-situ observations \cite{Barani2019,Barani2022}. }As radiation belt modelling improves, it becomes increasingly important to understand how all of these sources of uncertainty impact our final output to be able to analyse ensembles.

As uncertainty becomes increasingly important in radiation belt modelling, methods to account for uncertainty such as data assimilation and statistical methods are becoming the state-of-the-art method for modeling radiation belts\new{\ \cite{Koller2007,BourdarieMaget2012,Kellerman2014}}, while complex systems approaches are being applied to understand the underlying physics \cite{Balasis2023}. \cite{Morley2019} states that the field of space physics needs ensembles and methods to deal with and analyse ensembles, in order to manage uncertainty in parameterisations and in various parts of numerical space weather prediction. Ensembles, and other probabilistic \new{or statistical} methods, are already becoming the norm \cite{Hua2023,Watt2022,Sarma2020,Shprits2013}.

Satellite operators and national meteorological organisations are increasingly compelled to include space weather forecasting as part of their services, including the radiation environment. Geomagnetically trapped particles in the radiation belts have been established as a hazard to spacecraft due to processes such as surface charging, deep dielectric charging and single upset events \cite{Hands2018,Mateo-Velez2018}. 
 The loss of services such as the internet would cause severe socio-economic issues such as losses to navigation, finances (including card-based payments) and tracking/allocation of emergency services and commercial aircraft. 
These economic motivations for adapting modelling this fascinating system have also encouraged the space plasma physics community to adopt techniques mastered by meteorologists. However, the domain dependence of the diffusion coefficient means that using and analysing ensemble members is not simple, rendering ensembles less meaningful than desired.

To understand ensemble models for radial diffusion (and for radiation belt modelling more generally) we need to understand how simple model changes and initial and boundary conditions change the outputs, before we include variability to represent uncertainty in physical conditions.


\begin{notestocoauthors}
    {\color{purple}I don't know where this goes. I t was part of the mproject motivation.

Ideally, another output of an ensemble model would be a range of timescales for radial diffusion; although we understand many of the component processes, a timescale for radial diffusion is not something that has yet been clearly answered. Diffusion coefficients are often given in units of $days^{-1}$ but this does not explain the overall timescale when $D_{LL}$ can vary by orders of magnitude in time and space.}
    
\end{notestocoauthors}

\section{Background: Radial Diffusion Modelling in Earth's Radiation Belts}\label{sec:Background}

Radiation belt modelling is typically done using three adiabatic invariants; three quantities associated with the periodic motions of trapped, highly charged particles in Earth's magnetosphere. At relatively low energies (e.g. electrons of 1-100s of keV), particles are better described using the ring current paradigm. At higher energies (e.g. relativistic electrons $\geq 0.5$ MeV) 
we use the geomagnetically trapped radiation belt description \cite{RoedererZhangBook,WaltBook}. Periodic motions of particles in a slowly changing conservative field (such as electromagnetic fields) correspond to conserved quantities; adiabatic invariants. The first adiabatic invariant, \old{$\mu$}\new{$\mu_M$}, corresponds to the magnetic moment as particles gyrate in an electromagnetic field. The second invariant corresponds to momentum along the bounce path, as the gyrating particle travels up and down the field line, reflected by the stronger magnetic field ``magnetic bottle"). These motions are on scales of microseconds and seconds, respectively. Under the adiabatic approximation, if the system is changing slowly, then these quantities are conserved. On the other hand, if the system is perturbed on timescales comparable to these motions, the quantities \old{$\mu$}\new{$\mu_M$} or $J$ would not be conserved. On a much longer timescale, the periodic motion of electrons drifting around the Earth corresponds to the third adiabatic invariant $\Phi$, the magnetic flux enclosed by this drift path. This can be written as $\Phi = 2 \pi B_0 \frac{R_E^3}{R_0}$ for particles remaining at the equator, with $R_0$ the equatorial distance, $B_0$ the magnetic field strength at the Earth's surface and $R_E$ the radius of the Earth.  One can therefore model the radiation belts using these conserved quantities, by tracking the distribution of particles in this adiabatic invariant space and assuming that the change in this space can be modelled using diffusion - where diffusion between these occurs when electromagnetic perturbations occur on the temporal and spatial scales of the corresponding periodic motions. Diffusion is described using diffusion coefficients $D_{\alpha \beta}$. Typically, cross-terms corresponding to the third adiabatic invariant are considered to be zero; on the whole, changes to the thrid adiabatic invariant ("radial diffusion") is considered to be on a separate timescale and can therefore be approximated alone (e.g. \cite{Reeves2012}). The full diffusion model of the radiation belts has been used in a number of places, e.g. \cite{BeutierBoscher1995,Glauert2014,SubbotinShprits2009,Reeves2012}.

For the third adiabatic invariant, one can equivalently consider the drift shells instead of the magnetic flux conserved by the drift orbit. Since the drift shell is uniquely defined by its intersection with the equatorial magnetic field, it is often easier and more intuitive to parameterise the third adiabatic invariant using some form of $L^*$ parameter, which roughly translates to which nested drift shell a given electron is confined to. (From this point on we will refer to $L$-parameters, rather than $L^*$, for simplicity in notation). The drift shells a given $L$ correspond to will depend on the specific magnetic field model used, and multiple methods of defining $L$ exist depending on both the magnetic field and the approximations used to specify the drift shells (see e.g. \cite{Thompson2021,LejosneAlbert2023,WaltBook,RoedererZhangBook}). If one considers the $L$ parameter as a proxy for the relative radius of each drift shell, it becomes clear why diffusion across the third adiabatic invariant is called "radial diffusion": particles diffusing to different $L$ values are diffusing onto drift shells closer to, or further from the Earth. 
%
On the whole, radial diffusion is related to the large scale movement of particles towards and away from the Earth, acting to reduce gradients in the phase space density profile. The inward diffusion also corresponds to an increase in energy, if the first adiabatic invariant is conserved. Meanwhile, smaller scale processes breaking the first and second adiabatic invariants \new{can\ }result in local acceleration of particles (e.g. \cite{GreenKivelson2004,Reeves2013}).

The diffusion coefficient $D_{LL}$ contains the combined effect of electromagnetic perturbations on timescales corresponding to the electron drift. These perturbations consequently break the third adiabatic invariant, causing the diffusion to nearby $\Phi$ (or, equivalently, $L$). A comprehensive review of the current state of knowledge of radial diffusion coefficients can be found in \cite{LejosneKollman2020}. The first attempts to quantify this into a diffusion coefficient $D_{LL}$ assumed that such perturbations were stochastic; small scale continual ripples in the magnetosphere. The contributions from magnetic and electric potential perturbations were considered separately, using asymmetric and symmetric perturbations from a simple magnetic field model \cite{Falthammar1968}. Unfortunately these theoretical diffusion coefficients are problematic to apply; the magnetosphere is significantly more dynamic, rendering these assumptions invalid, while in practice one cannot observe these quantities to estimate them more accurately. There exists a gap between the theory and application of radial diffusion; accurate diffusion coefficients would require knowledge of the entire magnetosphere at each timestep to be able to calculate electron drift paths. Theoretical approaches must use magnetic field models and tend to focus on determining the validity of the underlying assumptions in order to derive a more appropriate method of calculating diffusion coefficients \cite{Lejosne2019,OsmaneLejosne2021,LejosneAlbert2023}. Any estimations of $D_{LL}$ used in modelling must make a significant number of approximations, most often based on the techniques in \cite{Brautigam2005,Fei2006} (particularly operational models). 

\cite{Fei2006} derives the diffusion coefficients for a particular storm using a formalism based on \cite{Falthammar1968}: for a given magnetic field model, find the deviation $\frac{dL}{dt}$ from drift contours due to azimuthally symmetric and asymmetric electromagnetic perturbations, and from this find $\left<(\Delta L)^2 \right>$ and hence $D_{LL}.$ \cite{Fei2006} uses a compressed dipole magnetic field, using an asymmetry factor $\Delta B$. \cite{Fei2006} also splits the diffusion coefficients into diffusion due to magnetic and electric perturbations, rather than into perturbations from induced electric and electric potential fields. \cite{Fei2006} used the larger component of the two to avoid counting the effect of the same perturbation twice, but the component which was dominant during their case study is not always the strongest one \cite{Olifer2019,Sandhu2021b}. In many subsequent studies, these components are simply added together to estimate $D_{LL}$. This makes the diffusion coefficients much easier to find and apply, but the resultant double-counting of perturbations means the final values could be off by around a factor of 2 \cite{Lejosne2019}. \new{However, uncertainty in $D_{LL}$ spans at least one and sometimes up to four orders of magnitude, e.g. across similar time periods} \cite{Sandhu2021b}, \new{between observations and models} \cite[Figure 6]{Olifer2019}, \new{and across different models at the same Kp and $L$} \cite[Figure 6]{Huang2010b}, \cite[Figure 4]{Liu2016}.\old{Given that the uncertainty in $D_{LL}$ across different methods in similar situations spans orders of magnitudes cite{Huang2010b,Liu2016,Olifer2019,Sandhu2021b}, this}\new{\ Given this uncertainty, and also the uncertainty in the relative size of induced electric and electric potential components,} double-counting is considered an acceptable compromise. Other choices made by \cite{Fei2006} based on properties of their storm-time magnetosphere are also replicated in subsequent methods, simply to have any ability to model the radiation belts at all. This may be why so many of these methods perform similarly - well on average, but with little specificity. 
A list of sources of uncertainty in $D_{LL}$ can be found in \cite{Bentley2019}.


\begin{notestocoauthors}
    {\color{purple}not yet integrated:
%
%
%
%
%
%
}
\end{notestocoauthors}

\section{Research Goals}\label{sec:goals}
Here we pause to explicitly identify research goals. Having a separate section for this allows us to report back on unfruitful research avenues and to directly label our goals to conclusion in the discussion, making the logic easy to navigate across several key results. No research paper has only a single important result and we hope this will bring out subtleties that may otherwise be missed.

We split these research goals up into: 
\begin{enumerate}
    \item \textbf{Primary goals} (research questions identified when scoping out the the project and applying for funding)
    \item \textbf{Secondary goals} (research questions that arose when developing our methodology)
    \item \textbf{Additional goals} (further questions that arose as part of the analysis which we realised can answer as part of this work)
\end{enumerate}
We choose to present our work in this way as it reflects our cy\new{c}lic methodology far more accurately, where we constantly revisited concepts and experiments until we reached a coherent structure explaining our results. 

\subsection{Primary Goals}
\begin{enumerate}
    \item[\textbf{[P1]}] Benchmarking for ensembles. 
    \begin{enumerate}
        \item How does one analyse ensembles for radiation belt modelling? How can we quantify differences between ensemble runs?
        \item Do changes in initial condition affect the differences between ensemble members? If so, which aspects of the initial condition have the strongest impacts?
        \item Do changes in model settings (such as $L_{outer}$) affect the differences between ensemble members?
    \end{enumerate}
    \item[\textbf{[P2]}] What is the timescale of radial diffusion? 
    \begin{enumerate}
        \item What is a useful definition of ``timescale"?
        \item What is the general radial diffusion timescale?
        \item Does timescale vary with initial conditions?
    \end{enumerate}
\end{enumerate}

\subsection{Secondary Goals}

\begin{enumerate}
    \item[\textbf{[S1]}] What do we do if we are not using a data-driven outer boundary? 
    \begin{enumerate}
        \item What boundary conditions would be physical to use? (rather than what is convenient for our observations)
        \item Where there are multiple potentially physical boundary conditions, how do they affect the timescale and evolution of radial diffusion?
        \item \old{What conditions should we use in future?}\new{What boundary conditions can we use in practice to balance physical boundaries (S1a) with observational and operational modelling constraints?}
        \item How does the choice of outer boundary condition and location interact with the current uncertainty on the outer boundary of radiation belt models?
        \item How does the outer boundary location interact with the increased diffusion coefficient at high $L$?
    \end{enumerate}
    \item[\textbf{[S2]}] What happens to an enhancement from local acceleration under radial diffusion (rather than just inwards diffusion of the ``background'' distribution, i.e. a source at high $L$?)
    \item[\textbf{[S3]}] How can time to monotonicity/ morphology of the particle distribution be used to analyse ensemble members? 
    \item[\textbf{[S4]}] Do we need to include loss \new{from precipitation via pitch-angle scattering} to represent radial diffusion?
\end{enumerate}

\subsection{Additional Goals}
\begin{enumerate}
    \item[\textbf{[A1]}] What analytical methods can be adapted to understand radial diffusion?
    \item[\textbf{[A2]}] How important are \new{PSD} gradients vs the $L$-dependence of $D_{LL}$?
    \item[\textbf{[A3]}] Is diffusion limited by the smallest value of $D_{LL}$ in the domain, i.e. the diffusion coefficient at lower $L$?
\end{enumerate}
\begin{notestocoauthors} 
\begin{enumerate}
        {\color{purple}
    \item[\textbf{[A0]}] \st{Stability maths for time and space varying Crank-Nicolson}
    \item[\textbf{[A4]}] \st{How do our measures relate to the energy of radially diffused particles?}
    }
\end{enumerate}
\end{notestocoauthors}

In the \cref{sec:Discussion} we will discuss our findings for each of these questions and outline future questions that are unanswered.

\begin{notestocoauthors}
{\color{purple} Are there more goals, are this actually what we're asking? S3b is simialr to P1b, there is nothing specific about N, E even in Additional Goals. Nothing about domain dependence - I suggest you add this as A4.
}
\end{notestocoauthors}

\section{Methods}
\label{sec:Methods}

In this section we will outline the scheme used for the numerical experiments in our ensemble and the metrics we use to analyse our experiments. We choose not to compare variability across the final phase space density distributions after a set time period, but to compare how several useful quantities vary across ensemble members.  We use a proxy (time to monotonicity, $t_m$) that indicates when radial diffusion has finished changing the shape of the distribution, and mass-like and energy-like techniques from analysis of dynamical systems to understand the ongoing evolution of the distribution. These tools allow us to see how radial diffusion is still contributing to radiation belt dynamics.

In \cref{sec:numexp} we discuss our equation for the initial condition, the parameters we will vary in our ensemble and the diffusion model we use. In \cref{sec:metricchoice} we will motivate time to monotonicity as a metric for timescale. In \cref{sec:mathsmethods} we derive the energy density-like and mass-like quantities we use to verify and interpret our simulations. 

Results from our investigation can be found in \cref{sec:results} and are brought into context with current understanding in \cref{sec:Discussion}.

\subsection{Numerical Experiments}
\label{sec:numexp}

\subsubsection{The Diffusion Model}
Although the full diffusion equation for the radiation belts models the phase space density (PSD) of electrons across all three adiabatic invariants, $M,J,\Phi$, the vastly longer timescales of drift motion means that one can separate out radial diffusion. We simulate radial diffusion following \cite{Thompson2020}. An idealised model allows us to examine how variation in initial conditions affect the final distributions, due solely to our modelling (rather than, for example, time-varying coefficients or inputs). We solve the radial diffusion equation
\begin{equation}\label{eq:diffusion}
\frac{\partial f(M,J,\Phi)}{\partial t} = L^2 \frac{\partial}{\partial L} \left( \frac{D_{LL}}{L^2} \frac{\partial f(M,J,\Phi)}{\partial L} \right)\,,
\end{equation}
Two sets of diffusion coefficients are used in this study. Where possible, the diffusion coefficient $D_{LL}$ is taken from \cite{Ozeke2012,Ozeke2014}:
\begin{align}
D_{LL}^E &= 2 . 16 \times 10^{-8} L^6 10^{0.217L + 0.461 Kp} \notag\\
D^B_{LL} &= 6.62 \times 10^{-13}L^810^{-0.0327L-0.108{Kp}^2 + 0.499 Kp} \label{eq:OzekeDLL}
\end{align}
in units of days$^{-1}$\new{, where $D^E_{LL}$ and $D^B_{LL}$ are the partial diffusion coefficients driven by electric and magnetic perturbations}. We scale these to units of seconds and take\new{ the approximation} $D_{LL} = D^B_{LL} + D^E_{LL}$. \cite{Ozeke2014} is calculated for a dipole magnetic field model using the symmetric components (equations 6 and 7 from \cite{Fei2006}). The asymmetric components disappear due to the dipole model; it is not clear how practical this is for radial diffusion as technically, symmetric perturbations existing for timescales significantly longer than a drift period should affect all electrons equally on a given drift orbit and therefore not break the third adiabatic invariant. Nevertheless, these diffusion coefficients are widely applied in practice due to the simple expression \cref{eq:OzekeDLL}, which was constructed with large numbers of observations and can reflect the changing dynamics of the magnetosphere in time by varying the geomagnetic activity index $Kp$, which takes values from 0 ("quiet") to 9 ("extremely active") \cite{Matzka2021}. In this work, we use the simplest method to define the drift shells of radiation belt electrons: equatorial $L$, where $L = 2\pi\Phi R_E^3/B_E$ in a dipole magnetic field. This suits both our idealised model and the diffusion coefficients used, which were calculated using a dipole model. Additionally, this $D_{LL}$ model performs similarly to the other most frequently used $D_{LL}$ \cite{Brautigam2005,Drozdov2021,Murphy2023}(i.e. it has similar order of magnitude), and enables comparisons with the variability study of \cite{Thompson2020}. Consistency here is important as we unpick the properties of radial diffusion; although in quiet times current models all perform similarly, it has been shown that during moderate geomagnetic storms, there is more variability between $D_{LL}$ estimation methods at a single event than from a single method, between two events \cite{Silva2022}. This model is also simple, similarly constructed to other empirical diffusion coefficient models and therefore should give us useful insight into the uncertainty from initial conditions. 

The $D_{LL}$ above reflects observed conditions, but as an empirical fitted function this $D_{LL}$ is not ideal for a mathematical analysis. Alternatives can be found by examining the derivation of \cref{eq:OzekeDLL}. The \cite{Ozeke2012} $D_{LL}$ were made by including electromagnetic perturbations $E_{total}, B_{total}$ across time and space in \cref{eq:Ozeke-general}
\begin{align}
D^E_{LL} &= \frac{1}{8 B_E^2 R_E^2}L^6 E_{total}(L,I) \\
D^B_{LL} &= \frac{L^8 4 \pi B_{total}(L,I)}{9 \times 8 B_E}, \label{eq:Ozeke-general}
\end{align}
and parameterising these using $L$ and $Kp$. In this analysis we are not interested in incorporating the observed electromagnetic perturbations across time and space, but are focusing on the system-scale behaviour. So we can base our $D_{LL}$ expression on \cref{eq:Ozeke-general}. Using either $n=6$ or $n=8,$  we can roughly say
\begin{equation}\label{eq:LnDLL}
D_{LL} = D_0 L^n,  
\end{equation}
which we will use to see the macro-scale movements of energy and mass in the system. To make our results comparable, we set $D_0 = D^E_{LL}(L=5,Kp=4)$ using \cref{eq:OzekeDLL} (so that $D_0 = 1.834 \times 10^{-5} days^{-1}$). This second model of $D_{LL}$ is needed to make the analytic approach in \cref{sec:mathsmethods} tractable.

\subsubsection{Diffusion model with loss}
Following initial experiments, we investigated whether our simple model required additional terms to validly represent the physics of the outer radiation belt. An important loss mechanism in the outer radiation belt is pitch-angle scattering (e.g. \cite{Lyons1972}), where the interaction between electromagnetic waves and electrons results in changes to the velocity vector relative to the magnetic field direction. In the relatively dense region of the plasmasphere, observed plasma and whistler-mode wave conditions are such that pitch-angle scattering is enhanced for high-energy electrons (e.g. \cite{Li2022})

Experiments were run twice, once without and then with loss from pitch-angle scattering included. While more sophisticated parameterisations for the electron lifetime exist (e.g. \cite{Orlova2016}) we require a version with very few parameters, and detail is unimportant as we are mostly interested in order-of-magnitude comparisons. Therefore, we use the simple electron lifetime from \cite{Shprits2007}
\begin{equation}
\tau \approx 1.2 E^2 L^{-1}    
\end{equation}
inside the plasmapause, measured in days${}^{-1}$ and with $E$ measured in MeV. We require $(\mu, J)$ to be constant. We choose our constant $\mu, J$  to be equal to that of a 2 MeV electron at $L=5$. Subsequently, the diffusion equation with loss is now
\begin{equation}\label{eq:diffusion+loss}
\frac{\partial f}{\partial t} = L^2 \left( \frac{D_{LL}}{L^2} \frac{\partial f}{\partial L} \right) - \mathcal{L} f\,,
\end{equation}
where
\begin{equation}
    \mathcal{L}=
    \begin{cases}
      \frac{1}{\tau}, & \text{if}\ L \leq L_p \\
      0, & \text{otherwise}
    \end{cases}
\end{equation}
for $L_p$ the edge of the plasmapause. This simple model is suitable for our investigation of idealised situations, with an additional parameter to explore (plasmapause location). In other experiments the default plasmapause is at $L=5$.

\subsubsection{Details of the Numerical Scheme}

Within this study, we use a modified Crank-Nicolson second order scheme, which has demonstrable success at numerically simulating the radial diffusion equation \cite{Thompson2020,Welling2012}, explicitly given by
%
\begin{widetext}
\begin{align}
\frac{f^{n+1}_j - f^n_j}{\Delta t} = \frac{L_j^2}{2} &\frac{1}{2 (\Delta L)^2} \left[ \frac{D_{j+\frac{1}{2}}^{n+\frac{1}{2}}}{(L _j + 0.5 \Delta L)^2} \left( f^n_{j+1} - f_j^n\right) - \frac{D_{j- \frac{1}{2}}^{n +\frac{1}{2}}}{(L_j - 0.5 \Delta L)^2} \left(f_{j-1}^n + f^n_j \right) \right. \notag \\ 
& \left.+   \frac{D_{j+\frac{1}{2}}^{n+\frac{1}{2}}}{(L_j+0.5 \Delta L)^2} \left( f_{j+1}^{n+1} - f_j^{n+1} \right) - \frac{D_{j-\frac{1}{2}}^{n+\frac{1}{2}}}{(L_j - 0.5 \Delta L)^2} \left( f_{j-1}^{n+1} + f_j^{n+1} \right)\right]\label{eq:CNscheme}\,. 
\end{align}
\end{widetext}
\cite{Tadjeran2007} shows that this scheme is unconditionally stable in the case where diffusion coefficients vary in time, but does not consider where the diffusion coefficients vary in space, as the coefficient matrix is then no longer symmetric. However, verification for our model can be found in the supplementary materials of \cite{Thompson2020} in addition to the initial SpacePy verification of \cite{Welling2012}.\new{ The timestep and spatial resolution of the simulations are $1$s and $0.1L$ respectively.}

\subsubsection{Initial Conditions}

\begin{figure}[h]
    \centering
    \includegraphics[width=\linewidth]{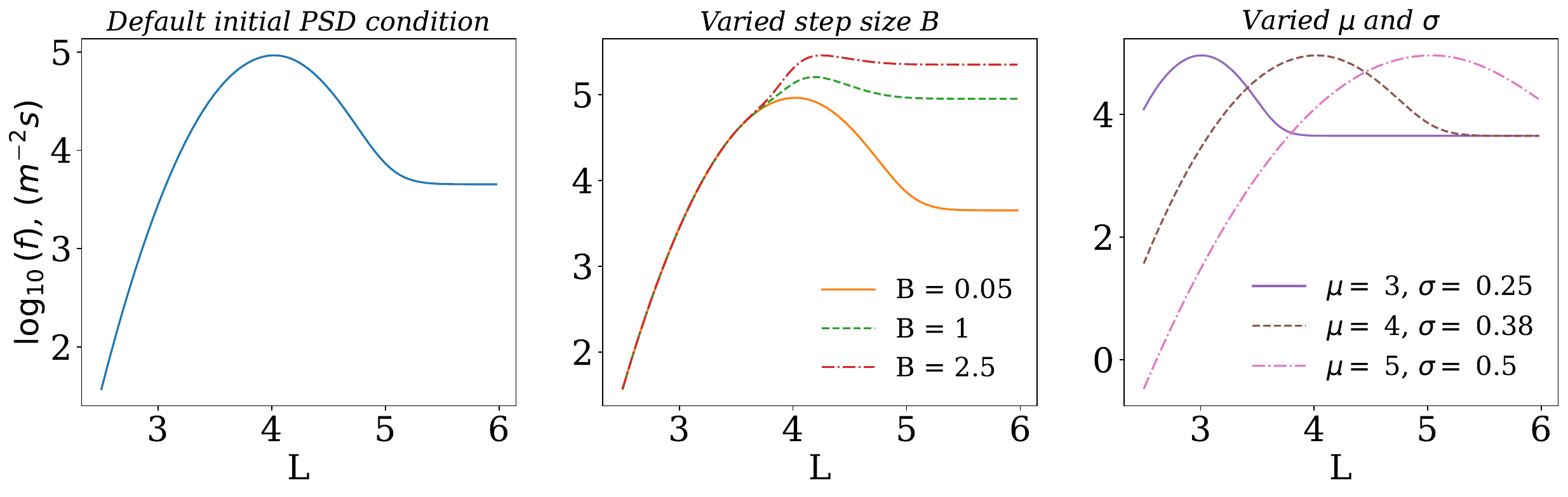}
    \caption{Possible configurations of the initial phase space density distribution function $f$. Combinations shown here are the default initial distribution (a) with $A = 9 \times 10^4,B = 0.05, \mu = 4, \sigma = 0.38, \gamma = 5$, (b) varied values of $B$ and (c) varied values of $\mu$ and $\sigma$.}
    \label{fig:IC}
\end{figure}

We characterise phase space density across drift shells, uniquely defined at the magnetic equator by $L$, using a distribution function $f$ \cite{Thompson2020}
\begin{eqnarray}
f(M,J,\Phi; t=0) &=& A \exp \left( - \frac{(L-\mu)^2}{2 \sigma^2}  \right) \nonumber \\ && + \frac{1}{2} A B \left[ erf(\gamma (L- \mu)) +1 \right] \label{eq:IC} 
\end{eqnarray}
This reflects typical phase space densities using a peak and step, which represents a state where inward radial diffusion has been occurring for some time (the step) and an enhancement of locally energised particles (the Gaussian).\new{\ These distributions were chose to reflect the phase space density dostributions observed in \cite{Boyd2018}.} Individual parameters are
\begin{itemize}
    \item $A$ amplitude
    \item $B$ step size (strength of error function)
    \item $\sigma$ width of density peak
    \item $\mu$ location of phase space density peak in $L$-space
\end{itemize}
Demonstrations of these initial settings can be found in \cref{fig:IC}. \new{Note that $B=0$ would result in an initial PSD distribution of only a Gaussian, while increasing $B$ represents a decreasing difference between height of the peak and height of the step. $B=1$ results in a step of comparable size to the enhancement, where the PSD at high $L$ is the same or greater than the pre-peak PSD. Although at $B=1$  the plateau asymptotes to the height of the Gaussian, since the two terms are summed together this is not a flat plateau. }Our default settings are $A = 9 \times 10^4,B = 0.05, \mu = 4, \sigma = 0.38, \gamma = 5$ and \old{$L_{outer}= 6.5.$}\new{$L_{outer}=6.$} The inner boundary is always set at $L=2.5$. We vary the location of the outer boundary ($L_{outer}$) to reflect the fact that radial diffusion models often have different $L_{outer}$, and to investigate the impact of different outer boundary locations when we know that diffusion varies in space as well as time.

\subsubsection{Outer Boundary Condition}
The model in \cite{Thompson2020} uses a constant value (Dirichlet) inner boundary to characterise the inner edge of the radiation belt, where particles are lost to the atmosphere. Choices of the outer boundary are less clear. Physically, one expects the PSD to be smooth across the boundary, hence a Neumann (constant gradient/zero flux) boundary may be appropriate. By default, the model uses a constant gradient (Neumann) outer boundary, set at zero, which matches the physics of a slow injection from high-L as represented in our initial condition. However, a Dirichlet boundary allows the modeller to input PSD values, and indeed large-scale models tend to interpolate time-varying outer boundary values from available data. Neither of these methods are designed to use the true edge of the outer radiation belt, which varies considerably when an outer edge is distinct enough to observe at all, \cite{Bloch2021}. Since both methods should be physically appropriate for the underlying plasma, in our results we investigate the impact of both choices of outer boundary condition. We will review this outer boundary in the discussion, following our results [S1a].

\subsection{Metric Requirements to Compare Ensemble Members: Time to Monotonicity}\label{sec:metricchoice}
Ensemble modelling for predictive purposes use the variability across model runs in the given ensemble, using a given error metric or loss function. We are also interested in examining timescale of radial diffusion, for which we need a quantitative measure.

However, error metrics are not an ideal tool for comparing the evolution of phase space density. They don't tell us about the evolution of the system state, or properties of that state we are interested in. Therefore one of the secondary goals of this work was to select and investigate potential metrics. Initially, simple error metrics such as mean square error (MSE) were selected. MSE would quantify a scalar difference  between distributions, which would be useful for analysis of variation and uncertainty across ensembles. However, the significant variation in scales covered in this problem make it difficult to generalise or compare different cases. We found this regardless of using linear- or log-based scales; thresholds usually would need to be specified for when radial diffusion was ``finished enough" or for when two distributions were ``similar enough", and the results became dependent on that choice of threshold. Instead, our experiments were analysed with a property that captured the physics of the system we were interested in: time to monotonicity, $t_m$. This choice of metric is motivated below. 
Requirements of the metric used for analysis are [A1]:

\begin{notestocoauthors}
{\color{purple}

    I admit it. We did not actually try the log-accuracy ratio, \cite{Morley2018}. It did not exist when we wrote this. If anyone has any thoughts on what we should add/ take away, let me know.
    }
\end{notestocoauthors}

\begin{itemize}
    \item \textbf{Robustness}. The metric used must be insensitive to any thresholds used. It must therefore be scale independent (i.e. work across multiple orders of magnitude, because of Kp dependence)
    \item \textbf{Interpretable} The metric must aid in understanding the system.
    \item \textbf{Radiation belt system specific.} The metric must be related to radiation belt modelling; it should provide insight into either the system state, or specific properties related to physical processes.
    \item \textbf{Time-series informative.} The metric must enable analysis of the evolution of system.
\end{itemize}

Initially, potential measures were tested, such as the decreasing MSE between distributions at each timestep, the evolution of maximum gradients, the total area and the proportional change in amplitude. All these required an arbitrary threshold to determine when a given experiment was “finished”, the choice of which strongly impacted the time until distributions became similar, particularly for larger Kp. For example, MSE-based metrics varied by orders of magnitude depending on several model choices. No MSE-based metric could be found that worked robustly. Furthermore, many of these metrics ended up being dominated by the inner boundary condition, whereas we wanted to know about the evolution of the entire phase space density distribution.

\begin{figure}
    \centering
    \includegraphics[width=\linewidth]{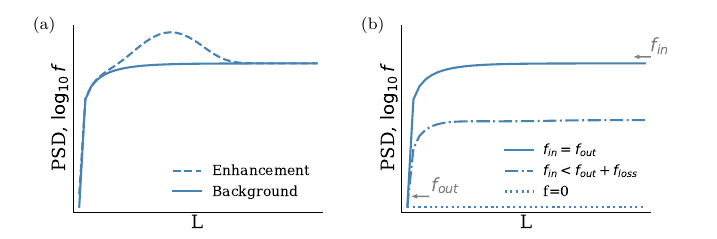}
    \caption{Example phase space density (PSD) distributions. Panel (a) shows the background, ``quiet" distribution in solid black lines, which is monotonic, plus a non-monotonic enhancement from local acceleration in dashed lines. Panel (b) shows potential long term (monotonic) PSD profiles: a return to the quiet profile (solid), a similar profile of reduced amplitude after loss from pitch-angle scattering (dot-dashed) and a zero profile (dotted).}
    \label{fig:diagram1and2}
\end{figure}

To find an appropriate metric of the how the ongoing radial diffusion affected the evolution of the system, we turned to properties of the PSD distribution under radial diffusion. \cref{fig:diagram1and2} (a) shows the phase space density profiles we expect on the timescale of radial diffusion. A quiet time, or background distribution, is shown in solid black lines: a low level, high $L$ source of particles feeds the system from constant substorms. Due to the $L$ dependence of the diffusion, the drop-off at low $L$ is quite sharp. In dashed lines, an enhancement of particles due to local acceleration is shown at an intermediate $L$. Following a single enhancement, we expect the distribution to gradually return to a monotonic state via radial diffusion \cite{Reeves2013,GreenKivelson2004}.

Therefore, since diffusion acts to even out gradients in the PSD, we are more interested in changes in the shape of the PSD distribution, than the total PSD (i.e. the integrated area under the curve). Hence we also include the following criteria:

\begin{itemize}
    \item \textbf{Insensitive to total particle population}. The metric must be insensitive to shifting the distribution up and down the y-axis.
\end{itemize}

We chose to work with \textbf{time to monotonicity}, $t_m$. When the distribution has become monotonic, the PSD distribution no longer has a peak for radial diffusion to smooth out. $t_m$ therefore indicates when the PSD distribution has stopped changing shape; when radial diffusion is no longer changing the properties of the radiation belts. $t_m$ is a proxy for whether radial diffusion is still significantly affecting the evolution of the particle distribution. The potential monotonic distributions are shown in  \cref{fig:diagram1and2} (b); the ``background" distribution which is nonzero but unchanging in time (i.e. the influx of particles balanced by constant movement of particles inwards), a shrinking version of this when more particles are lost than enter the domain, and finally a zero profile.

In the results section we use $t_m$ to explore how idealised radial diffusion models vary when changing initial settings. $t_m$ represents our physical expectations; our intuition that after a localised enhancement the PSD will eventually relax to monotonic distribution. We can test our expectations and how the changing parameters affect these expectations. See \cref{sec:Discussion} for evaluation of our chosen measure and for alternative approaches.

\subsection{Analytical Approach to Comparing Evolution of Ensemble Members}
\label{sec:mathsmethods}
To study the impacts of varying initial and boundary data, we will utilise tools from \new{t}he study of deterministic diffusive problems. In particular, we will monitor the number of particles and an energy-like quantity in our simulation domain, defined respectively by the following integrals [A1]:.
\begin{align}
\mathcal{N} &= \int  \frac{f}{L^2} dL\,, \label{eq:N}\\
\mathcal{E} &=  \int   \frac{f^2}{L^2} dL\,. \label{eq:E}
\end{align}
The former of these ($\mathcal{N}$) is the conventional integral of the distribution function with the appropriate Jacobian for the radial component of the co-ordinate system, due to the use of adiabatic invariant variables. 
The latter ($\mathcal{E}$) is unconventional in the study of radial diffusion as far as the authors are aware, but is closely related to the $L_2$ norm of the distribution function. \new{The $L_2$ norm is a positive-definite measure of a function that is typically used to understand the magnitude of a given function on a specified domain. For example, the $L_2$ norm of vectors can also be written as $|x|=\sqrt{\mathbf{x}^2}=\sqrt{x_1^2+x_2^+\ldots + x_k^2}$.}
Such an integral is a linchpin in the study of diffusion in other contexts~\cite{Strauss2007}.
We refer to this integral as an `energy' integral in a loose sense, as it is a positive definite quantity throughout the distribution functions evolution and is minimized precisely when the distribution function's evolution has ceased (i.e. when $f_t = 0$).
We will investigate how these quantities change in the system and use these changes to confirm or clarify the results from our ensemble runs analysed using $t_m$. 

In order to understand how these quantities change with evolving $f$ it will be useful to monitor the rate of change of these integrals over time. These rates can be computed explicitly as follows:
\begin{align}
\mathcal{N}_t &= \int f_t \frac{1}{L^2} dL = \int \left( \frac{D_{LL}}{L^2} f_L \right)_L dL = \left[ \frac{D_{LL}}{L^2} f_L \right]^{OB}_{IB} \label{eq:Nt}\\
&\notag \\
\mathcal{E}_t &= \int f f_t \frac{1}{L^2} dL = \int f \left( \frac{D_{LL}}{L^2} f_L \right)_L dL \notag \\ 
&= \left [ \frac{D_{LL}}{L^2} f f_L \right]^{OB}_{IB} - \int \frac{D_{LL}}{L^2} \left( f_L \right)^2 dL\,, \label{eq:Et}
\end{align}
where $OB, IB$ indicate evaluating the resulting function at the inner and outer boundaries respectively.
The rate of change of energy $\mathcal{E}_t$ is a useful diagnostic to determine whether the distribution function $f$ is approaching its equilibriated state, as this will be reflected by this rate of change approaching zero. What is clear from the final forms in (\cref{eq:Nt}) and (\cref{eq:Et}) is that the choice of the diffusion coefficient significantly impacts these rates and differing choices may either accelerate or arrest the dynamics as they approach monotonicity. Within this paper, we will restrict ourselves to a specific form of this coefficient, made explicit in the next section, but it is clear that alternative choices may impact the conclusions we draw from this study. Further, these rates are explicitly dependent on the boundary conditions chosen for the simulation as well as the size of the domain. We will return to dependence later when we discuss outcome of the numerical experiments we undertake as the key study of this paper.

As a final comment, the above rates of change generalise quite naturally when loss is included within the radiation belt monitoring. Recall that when loss is included, the radial diffusion equation assumes the form
\begin{equation*}
    f_t = L^2 \left( \frac{D_{LL}}{L^2} f_L \right)_L - \mathcal{L} f\,.
\end{equation*}
It then follows by repeating the analysis above that the loss-modified rates of change for the number and energy are given by
\begin{align*}
\mathcal{N}_t^{loss} &= \mathcal{N}_t -\int_{IB}^{L_p} \mathcal{L} \frac{f}{L^2} dL \\
&= \mathcal{N}_t -\int_{IB}^{L_p} \frac{1}{\tau(L)} \frac{f}{L^2} dL \label{eq:Ntloss}\\[3mm]
\mathcal{E}_t^{loss} &= \mathcal{E}_t -\int_{IB}^{L_p} \frac{1}{\tau(L)} \frac{f^2}{L^2} dL\,.
\end{align*}
As the new terms are positive definite, the loss effects cause a continual loss of energy as expected, with equilibrium only occurring when $f=0$ for all $L$, or when inward flux from the outer boundary equals the combined loss from the inner boundary and pitch angle scattering. Again, we would also expect particles to move towards lower $L$.

In this section we derive the terms that comprise the changing energy ($\mathcal{E}_t$) and mass ($\mathcal{N}_t$). The figures comparing these terms, evaluated explicitly for each of our boundary conditions, can be found in \cref{sec:results} where they best support our analysis.

Using our $D_{LL}$ from (\cref{eq:LnDLL}), we may decompose the contribution to the rate of change of number and energy respectively in the following way: 
\begin{align*}
\mathcal{N}_t &= \left[ D_0 L^{n-2} f_L \right]^{OB}_{IB} 
\\
&= -\mathcal{N}_t^{(1)} - \mathcal{N}_t^{(2)} \\
\mathcal{E}_t &= \left[ D_0 L^{n-2} f f_L\right]^{OB}_{IB} - \int D_0 L^{n-2} \left( f_L \right)^2 dL \\
&= -\mathcal{E}_t^{(1)} - \mathcal{E}_t^{(2)} - \mathcal{E}_t^{(3)} \,.
\end{align*}
i.e $\mathcal{N}_t$ has two terms and $\mathcal{E}_t$ has three. We can now evaluate the role of each term on the distribution function's evolution, and we start by discussing the effects of the boundary conditions on the number and energy.  It is clear that the number of particles in our system can only change based on the boundaries; whether this is a loss or addition will depend on the gradient at each boundary, and the magnitude of that change will be moderated by the location of that boundary owing to the fact that the diffusion coefficient depends on the spatial co-ordinate $L$. The ‘energy’ can not only change due to boundary effects (terms 1 and 2), but also due to an additional term $\mathcal{E}^{(3)}_t$ which is dynamical in origin. Terms 1 and 2 depend on the PSD, the gradient of the PSD at that boundary and the location of that boundary. Both of these can contribute to energy increases or decreases in the same way that the number density varies at the boundaries, but the third term informs us on how the energy is minimized due to effects of the distribution function on the interior of our domain. We discuss these effects and their implications for $f$ below.

The third term in $\mathcal{E}_t$ is one that will give us insight into how the dynamics of the distribution will evolve to minimise our energy. This term has an integral which depends on the square of the derivative in $L$. As this is a non-negative quantity, this ensures that the integral contributes to energy loss (as it is preceded by a negative sign) whilst gradients in the distribution function exists, up to a point where the contributions from the boundary conditions balance this out. This is an property identical to Cahn-Hilliard diffusive dynamics, where diffusive terms correspond to energetics that penalise the formation of (sharp) gradients~\cite{Kendon2001,Stratford2005}. This penalty is enhanced for gradients occurring at higher values of $L$, as their contribution to the integral will be increased by a positive power of $L$, suggesting that the distribution function can minimise this integral by moving gradients to lower values of $L$. Thus, $\mathcal{E}_t$ decreases towards a long-term solution through diffusion to reduce gradients $(f_L)^2$ and through the population moving towards lower $L$, and this movement of the population will be increased for higher powers $n$. 



\section{Results}\label{sec:results}
Our methodology was to begin with $t_m$, to find out how long it takes a given initial distribution to reach a monotonic state, and how this varies with initial conditions. We first compared $t_m$ with Kp for each parameter in \cref{eq:IC}. For $t_m$ we note that from \cref{eq:OzekeDLL}, we expect timescale to vary significantly with Kp. Kp is a proxy for strength of radial diffusion, even though $Kp > 6$ is unlikely. Since we expect time to monotonicity to depend on Kp we primarily use a heatmap for the ensemble used to investigate each parameter, demonstrating how $t_m$ varies with each (Kp, parameter) pair. To aid understanding, these results are also presented in an alternate format, where the $t_m$ for each Kp are plotted as a line. Each experiment ran for a week; model runs where monotonicity were not reached are left empty. 

Following our initial $t_m$ analysis, we then investigated any noteworthy results, for example by looking at the evolution of the phase space density of a specific simulation. In general, runs with a Neumann outer boundary condition (zero flux) are shown on the left, while the right hand column corresponds to a Dirichlet outer boundary condition (constant value)

Finally, we incorporated our analysis from \cref{sec:mathsmethods} for each parameter to understand and generalise any patterns we saw. Our $\mathcal{N,E}$ experiments are also run for a week, using $n=6$ in our diffusion coefficient. Just as for our $t_m$ analysis, we calculated $\mathcal{N,E}$ for each returned timestep (every six hours). 

\subsection{Results Part 1: Without loss rate}\label{sec:resultspart1}

Each parameter in the initial condition was systematically investigated. Selected results are presented in the main text in an order chosen to best convey our conclusions; the full set of individual $t_m,\mathcal{N}$ and $\mathcal{E}$ experimental figures can be consulted in the supplementary material, labelled as Figure SX.\new{ As a one dimensional simulation, the number of particles in a given phase space bin (PSD) is in units of $(m^{-2}s)^1$ (SI units, per (unit speed $\times$ unit length)). This can be scaled to any of the other units systems used for PSD elsewhere.}

\subsubsection{The difference between the two outer boundary conditions}
\begin{figure}[ht]
    \centering
%
    \includegraphics[width=\linewidth]{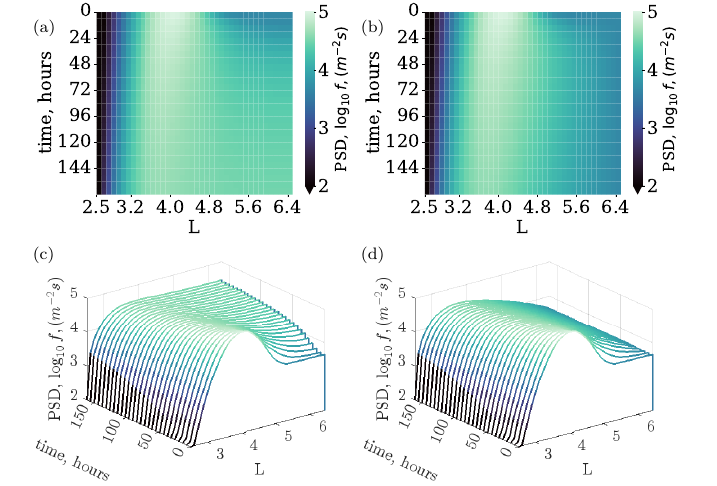}
    \caption{Phase space density over a week, with Neumann (constant gradient, left) and Dirichlet (constant value, right) outer boundaries.}
    \label{fig:PSDtime}
\end{figure}

Overall, it is clear from Figure S1 that generally, more runs with a Neumann (fixed gradient) outer boundary reach a monotonic state within a week\new{, compared to runs with a Dirichlet (fixed value) outer boundary}. We will investigate the two outer boundary conditions before examining the effect of each parameter in the initial condition. To understand the evolution of $f$ in these experiments, in \cref{fig:PSDtime} we show the phase space density for both Neumann and Dirichlet outer boundary conditions, with an outer boundary located at $L_{outer} = 6.5$. We use default settings for the initial PSD, a Kp of 4 and run for a week. Panels (a) and (b) show the heatmap, while waterfall plots are in panels (c) and (d).

\begin{figure}[ht]
    \centering
    \includegraphics[width=\linewidth]{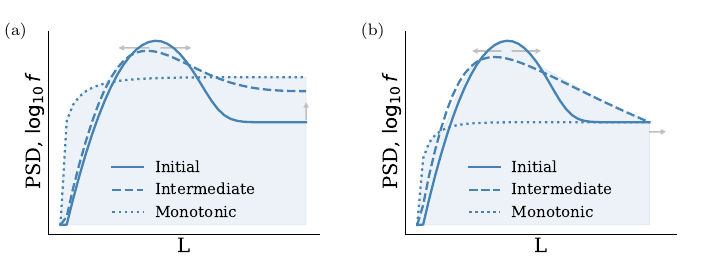}
    \caption{How phase space density profiles with Neumann and Dirichlet outer boundary conditions reach a monotonic state.(Left) Neumann runs reach monotonicity by the high-$L$ population increasing. The Dirichlet runs (right) reach monotonicity by material leaving the domain until the peak has diffused away. The point at which each reach monotonicity is very different (see monotonic states in dotted blue, and intermediate states in dashed blue.)}
    \label{fig:diagram3}
\end{figure}

In both experiments the peak remains high and moves inwards. However, there is a difference in the outer part of the simulation\new{, as demonstrated in \cref{fig:diagram3}}. The plateau becomes significant for the Neumann boundary but not for the Dirichlet boundary. This makes sense as the outer boundary value is able to rise for the Neumann case, reflecting outward radial diffusion. For Dirichlet runs the outer boundary value cannot rise, but particles can be lost. Note that the amplitude of the Neumann peak is still comparable to the Dirichlet case (4.55 and 4.52 $\log_{10}(PSD)$ at $L=3.9$ and $3.7$ respectively); it is the plateau that has changed. 
More Neumann experiments reach monotonicity as the plateau can rise instead of having to wait until the peak diffuses completely inwards. This inability for the high-$L$ (to the right of the peak) PSD to reach (positive) monotonicity independently of the left part is one reason why it takes longer for the Dirichlet runs to reach monotonicity. This corresponds to outward radial diffusion varying depending on the outer boundary condition [S1b].


For a more realistic outer boundary condition we need to consider these, along with the fact that we don't currently have a clear outer boundary location. We discuss all these factors together in \cref{sec:Louter_discussion}.

\begin{figure}[ht]
    \centering
    \includegraphics[width=\linewidth]{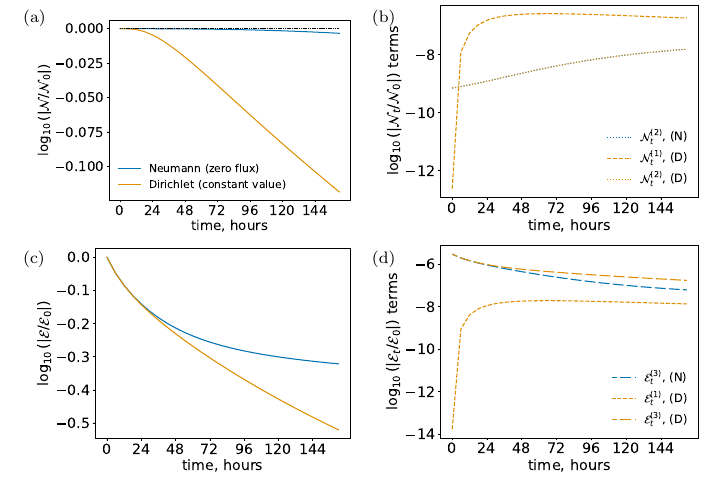}
    \caption{The value of our mass-like quantity, $\mathcal{N}$ and our energy density-like quantity, $\mathcal{E}$, across a week for both Neumann (blue\new{\ `(N)'}) and Dirichlet (orange\new{\ `(D)'}) outer boundary conditions. (a) shows how $\mathcal{N}$ evolves across the week; (b) shows how loss from the outer and inner boundaries contribute to a reduction in $\mathcal{N}$ ($\mathcal{N}_t^{(1)}$ and $\mathcal{N}_t^{(2)}$ respectively); (c) shows how $\mathcal{E}$ evolves across the week; (d) shows how loss from outer and inner boundaries and from the reconfiguration contribution to the reduction in $\mathcal{E}$ ($\mathcal{E}_t^{(1)},\mathcal{E}^{(2)}_t$ and $\mathcal{E}_t^{(3)}$ respectively).\new{\ Solid lines are used for $\mathcal{N}$ and $\mathcal{E}$ whilst different linestyles are used for components of $\mathcal{E}_t,\mathcal{N}_t$.} (a) and (b) are both normalised by $\mathcal{N}$ at time $t=0$ and (c) and (d) are normalised by $\mathcal{E}$ at time $t=0$.}
    \label{fig:NvsD_EN}
\end{figure}

We can also examine how the evolution of our mass- and energy-like densities varies with Neumann and Dirichlet outer boundary conditions. \cref{fig:NvsD_EN} shows $\mathcal{N,E}$ across the week, plus the components of $\mathcal{N}_t,\mathcal{E}_t$. Panels (b) and (d) show the absolute value of components on a log scale to make order of magnitude comparisons earlier; in rare cases in later analyses where the change becomes positive (i.e. a gain in mass or energy, rather than a loss), this is specified.

By definition, for Neumann experiments $\mathcal{N}_t^{(1)} = 0$, i.e. mass is only lost through the inner boundary. For the Dirichlet experiments it is clear that the outer boundary dominates loss, with $\mathcal{N}^{(1)}_t$ up to two orders of magnitude larger than $\mathcal{N}^{(2)}_t$, hence the number of particles decreases more rapidly. We find that less than 0.8\% of the original mass is lost with a Neumann outer boundary, while around 24\% of the mass is lost with a Dirichlet boundary [S1b].

Comparing the individual terms contributing to changes in mass \new{in \cref{fig:NvsD_EN} (b)}, we see that loss from the inner boundary \new{$\mathcal{N}^{(2)}$ }is the same across the week regardless of the outer boundary (and therefore independent of the different interior distribution as the system evolves). For inner boundary loss to be big enough to vary, one must run experiments with very large diffusion coefficients (e.g. using $Kp=9$ in \cref{eq:OzekeDLL}) or for a much longer time.

The results for our $L_2$ norm $\mathcal{E}$ are somewhat counterintuitive. In total we know that experiments with a Neumann outer boundary can eventually reach a lower-$\mathcal{E}$ state (zero everywhere) than experiments with a fixed outer boundary value, where the minimum-energy state will have the same fixed value at $f|_{OB}$ as the initial condition. However, we find that Neumann simulations appear to be reaching a limiting state, where $\mathcal{E}_t$ is increasingly smaller and $\mathcal{E}$ relatively unchanged. 

For a Neumann outer boundary, $\mathcal{E}_t = -\mathcal{E}_t^{(3)}$. While the Dirichlet outer boundary can contribute to changing energy ($\mathcal{E}_t = -\mathcal{E}_t^{(1)}  -\mathcal{E}_t^{(3)}$ ) we can see from \cref{fig:NvsD_EN}(d) that the dominant mechanism for energy loss is mostly from the reconfiguration term $\mathcal{E}^{(3)}$. However, this term accounts for more energy loss when the outer boundary is Dirichlet rather than Neumann, because there are more steeper gradients when the outer boundary is fixed. Since $\mathcal{E}_t$ depends on gradients ($f_L$), $\mathcal{E}_t$  is larger for Dirichlet runs as there are gradients both sides of the peak, rahter than just to a plateau. For Neumann experiments, the gradients rapidly flatten into a plateau at higher $L$. Remember that our equation for $\mathcal{E}$ has a factor of $\frac{1}{L^2}$ in it: the same PSD at a higher $L$ contributes less to the norm, because it can be moved around more easily. Hence Neumann runs have a lower $\mathcal{E}_t^{(3)}$. Once the plateau has been reached, the only way to lose energy is for material to move down the gradient (and then out through the inner boundary). This process is slow and so $\mathcal{E}$ is effectively limited. On the other hand, the Dirichlet experiments are more effectively moving material to higher $L$, and then out of the domain. Having an outer boundary that allows flux in/out means that $L_2$ norm is reducing more quickly than when we allow the PSD at the outer boundary to change. Hence Neumann appears to be reaching a limiting state first, where $\mathcal{E}_t=0$ and $\mathcal{E}$ is unchanging, even though we know that if left forever, it can reach a lower-energy state than Dirichlet [S1b;S1e].

\subsubsection{Significant properties of the initial condition ($\mu,B$)}

The results of systematically investigating each parameter, using first time to monotonicity $t_m$ and then our quantities $\mathcal{N,E,N}_t,\mathcal{E}_t$, are shown here. Those properties we have considered to have a significant impact on time to monotonicity are presented in detail, whilst remaining properties are covered briefly in \cref{sec:minor} [P1b;S3a].

\textbf{Step size $B$:} The step size $B$ corresponds to a system where the phase space density is higher further out in the radiation belts; i.e. a situation where inwards radial diffusion from a distant source has already occurred. A higher step therefore corresponds to radiation belts that have more material before the enhancement occurs.

\begin{figure}[ht]
    \centering
    \includegraphics[width=\linewidth]{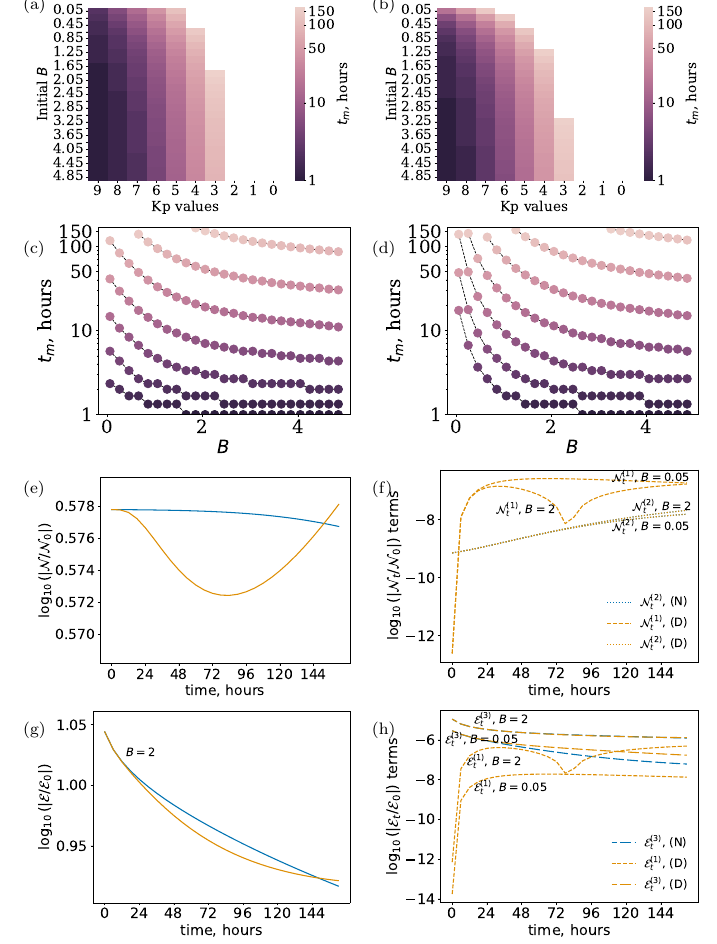}
    \caption{Selected results for the impact of increase in step size $B$ on the evolution of the system under radial diffusion. \old{Left: Neumann. Right: Dirichlet. }The top two rows show the change in time to monotonicity $t_m$ with $B$\old{,}\new{for Neumann (left) and Dirichlet (right) outer boundary conditions. (a) and (b) show the time to monotonicity as a colourmap with increasing $B$ and Kp. Panels (c) and (d) are an alternative view with the same colourscales, where each line corresponds to one Kp value - i.e. one row of the panel above.}\old{ whilst the}\new{ The} bottom panels show the changing mass-like and energy-like quantities $\mathcal{N}$\new{ (e)} \old{and }\new{, }$\mathcal{E}$\new{ (g) and the components of $\mathcal{N}_t$ and $\mathcal{E}_t$, (f) and (h) respectively. In the bottom panels, Neumann experiments are indicated by blue line and Dirchlet by orange. Solid linestyles indicate $\mathcal{E},\mathcal{N}$ whilst different linestyles indicate the components of $\mathcal{E}_t$ and $\mathcal{N}_t$}. $\mathcal{N, E}$ and $\mathcal{N}_t^{(1,2)},\mathcal{E}_t^{(1,2,3)}$ are normalised by the initial values of $\mathcal{N}$ and $\mathcal{E}$ respectively. All $\mathcal{N}_t, \mathcal{E}_t$ terms are shown as absolute values to enable a log scale.}
    \label{fig:B_par}
\end{figure}

\cref{fig:B_par} (a) and (b) show how time to monotonicity $t_m$ varies with both Kp and increasing step size.\new{ Blank values indicate monotonicity was not reached within a week, the length of the experiments.} When starting with a larger step size, we find that monotonicity is reached sooner for both outer boundary conditions. This behaviour is as expected as with an increased $B$, $f$ is already closer to a monotonic state. Again, more ensemble runs with Neumann outer boundary reach monotonicity. Looking at this information in the alternative format in panels (c) and (d), we see that for both outer boundary conditions, $t_m$ appears to increase exponentially for smaller step sizes.

Our $\mathcal{N,E}$ results show that the evolution is not necessarily straightforward, however. \cref{fig:B_par} (e) shows the total mass in $f$ across the week simulated, for the larger step size $B=2$ for both Neumann and Dirichlet outer boundary conditions (the comparison against the default value can be found in Figures S3 and S4). As expected, the total number of particles changes very little for a zero flux (Neumann) outer boundary. However, \old{when }the overall mass response\new{\ changes} when mass can flow across the outer boundary (Dirichlet experiments)\old{ we}\new{. We} see that for a larger step size $B=2$, the experiment starts to gain mass at around 80 hours. From the mass change terms $\mathcal{N}_t$ in \cref{fig:B_par}(f) this is clearly from the outer boundary, when the distribution drops below the fixed outer boundary point $f|_{OB}$. At this point the gradient $f_L|_{OB}$ will be\new{come} positive, and material will flow into the domain. \old{(The plain $\mathcal{N}_t,\mathcal{E}_t$ components can be found in the supplementary materials for confirmation that this becomes positive, rather than the absolute log values shown here). }Inner boundary loss varies with $B$ but but is again independent of the outer boundary condition.

The $L_2$ norm $\mathcal{E}$ is much higher for a larger step, and the norm reduces over several orders of magnitude (see \cref{fig:B_par}(g)). The Dirichlet run started out losing more energy than Neumann (as we also see using default settings above) but later in the week, energy loss drops off and the Neumann case has lower energy (and is therefore closer to a point where the dynamics have stopped changing). This is because there is an increase in the norm ($\mathcal{E}$) with the reversed outer boundary flow; however, the corresponding energy change term $\mathcal{E}_t^{(1)}$ reaches a comparable magnitude to the dominant reconfiguration term $\mathcal{E}^{(3)}_t$. Therefore the total change $\mathcal{E}_t$ for the Dirichlet case with $B=2$ becomes very small, while the Neumann case is still reducing in norm because the PSD can be reconfigured (diffused).

Physically, this means that with a higher step, we are finding that the Neumann case reaches a lower energy state by the end of the week than the Dirichlet experiment, unlike our default settings \cref{fig:NvsD_EN}. For the Dirichlet case, the constant outer boundary value is higher than the peak, allowing material to come in through the outer boundary. This experiment is in a state where the main dynamic is a constant churn of mass coming in and then being diffused to lower $L$ to reduce the entire distribution to a lower energy state. Physically, this corresponds to an infinite source at the outer boundary if we were to run this indefinitely. See \cref{sec:Discussion} for our overall conclusions on more suitable outer boundary settings.

\textbf{Enhancement location $\mu$:} The Gaussian in the first term of \cref{eq:IC} corresponds to an enhancement, and $\mu$ corresponds to the location of this enhancement in $L$.

\begin{figure}[ht]
    \centering
    \includegraphics[width=\linewidth]{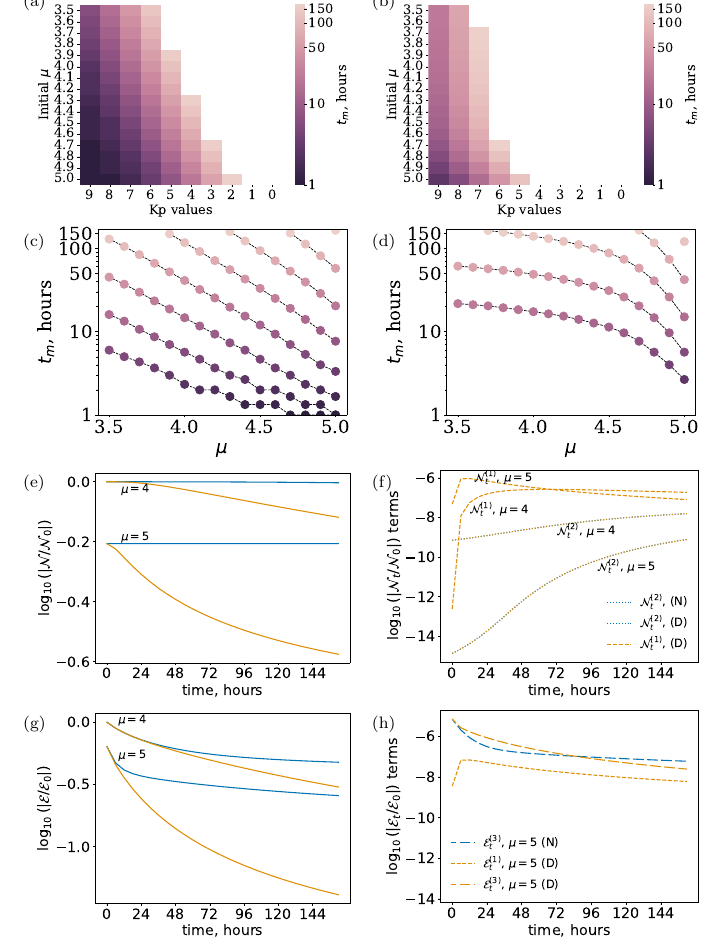}
    \caption{Selected results for the impact of increase in enhancement location $\mu$ on the evolution of the system under radial diffusion. As in \cref{fig:B_par}, the top two rows pertain to time to monotonicity $t_m$ and the bottom two to the mass-like quantity $\mathcal{N}$ and the $L_2$ norm $\mathcal{E}$. Neumann and Dirichlet outer boundary conditions are shown in the left and right columns respectively.}
    \label{fig:mu_tm}
\end{figure}

\begin{notestocoauthors}
    {\color{purple} Co-authors, when we have two $t_m$ and two $\mathcal{N,E}$ rows, do you prefer figures split up like in $\mu$, or merged like in $B$? }
\end{notestocoauthors}
Considering $t_m$ across a variety of enhancement locations in \cref{fig:mu_tm}, we find that monotonicity is reached more quickly for higher $\mu$ for both outer boundary conditions, although the shape of the dependence is seen to be very different in \cref{fig:mu_tm}(c) and (d). A sooner $t_m$ for higher $\mu$ makes sense as $D_{LL}$ will be larger at higher $L$, so when the peak is located at higher $L$, diffusion to flatten this peak happens more quickly. Again, far more runs reach monotonicity with a Neumann outer boundary.

\cref{fig:mu_tm}(a) and (b) show us the general relationship between each parameter and $t_m$, while (c) and (d) show us the specifics of each relationship. We see that the relationships between $\mu$ and $t_m$ for each Kp are quite different for the different outer boundary conditions. This disparity could be due to the different mechanism to the right of the enhancement; a zero flux outer boundary condition allows the plateau to rise and reach monotonicity to the right of the peak, while the same region with a constant value outer boundary cannot reach monotonicity until the enhancement has completely diffused, although material can be lost through the boundary.

The bottom two rows of \cref{fig:mu_tm} compare the evolution of $\mathcal{N,E}$ across the simulation for the default value of $\mu=4$ to a more distant peak at $\mu=5$. The total number of particles $\mathcal{N}$ is higher when $\mu$ is lower, which is as expected since the step extends further inwards. There is little change in mass with $\mu$ for Neumann runs, also as expected. There appears to be different amounts of mass lost in Dirichlet simulations, so we consider the outer and inner boundary particles losses $\mathcal{N}_t^{(1)}$ and $\mathcal{N}_t^{(2)}$ in (f). These are always negative (particles are only lost, not gained) but look quite different. The outer boundary loss dominates for both values of $\mu$. With a higher-L enhancement, the inner boundary loss is less. Therefore whilst the outer boundary loss is comparable, the higher the $\mu$, the more strongly that outer boundary loss dominates over the loss from the inner boundary. 
\cref{fig:mu_tm} (g) shows the evolution of $\mathcal{E}$. Lower values of $\mu$ (enhancements at lower $L$) actually result in higher $L_2$ norms, because you have more PSD total (for the same reason as the mass above) and more of this mass is at lower $L$. In both cases the Neumann experiments reach a configuration where $\mathcal{E}_t$ is very small and $\mathcal{E}$ stops changing. The Dirichlet experiment with a higher-L enhancement rapidly loses $\mathcal{E}$ but by the end of the week, is no longer changing much. In (h) we can see the terms of $\mathcal{E}_t$ for each experiment. For readibility, we show only the $\mathcal{E}_t$ terms for $\mu=5$ here. The reconfiguration energy loss ($\mathcal{E}_t^{(3)}$) evens out more quickly for Neumann experiment with $\mu = 5$ than with the standard initial condition, presumably because it is easier for the step to rise up and plateau in a monotonic state. For the Dirichlet experiment the reconfiguration term $\mathcal{E}_t^{(3)}$ dominates, but rapidly drops off until it is comparable with energy loss at the outer boundary.

\begin{notestocoauthors}
{\color{purple}Clare, Dan, this does not feel insightful. Can I something more interesting, in fewer words?}
\end{notestocoauthors}

We find that an initial distribution with more material at high $L$ (i.e. step size $B$) and with an enhancement at high $L$ (i.e. $\mu$) diffuse more quickly. $B$ and $\mu$ are the most significant initial conditions, yet the specific evolution of the system varies depending on interaction with boundary conditions [P1b].

\subsubsection{Minor properties of the initial condition ($A,\sigma$)}\label{sec:minor}

\textbf{Amplitude $A$:} We do not expect the amplitude of the initial condition to impact our time to monotonicity. This is because the radial diffusion equation is linear and so amplitude scalings can be factored out. Indeed, setting $f = A g(L,\mu,\sigma,t)$ one finds that
%
 \[   f_t = L^2 \left(\frac{D_{LL}}{L^2} f_L \right)_L \qquad \Rightarrow \qquad g_t = L^2 \left( \frac{D_{LL}}{L^2} \cdot g_L \right)_L\,,
 \]
%
which will have the same solutions as \cref{eq:diffusion}, which are independent of $A$. As monotonicity is a property of a given solution, the time to reach this monotonic solution is independent of $A$. This is verified in the supplementary materials (Figure S1 (a), (b)) where we observe that time to monotonicity varies with $Kp$ as expected and does not vary with initial amplitude. We note that although the energetic quantities $\mathcal{N,E}$ scale with $A$ and $A^2$ respectively, they evolve on the same timescales as \cref{fig:NvsD_EN} via similar arguments to the above.

\begin{figure}
    \centering
    \includegraphics[width=\linewidth]{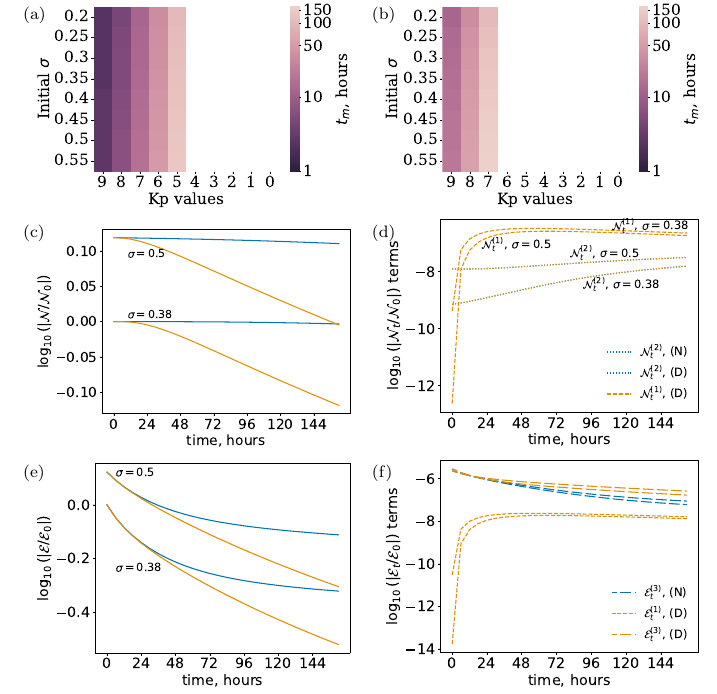}
    \caption{(Top row) time to monotonicity $t_m$ for enhancement width $\sigma$ vs Kp over a week, for Neumann (a) and Dirichlet (b). (Middle row) The mass-like quantity $\mathcal{N}$ for $\sigma=0.38$ and $0.5$, and the components of the time derivative $\mathcal{N}_t$. (Bottom row) Same as the middle row, but for the $L_2$ norm $\mathcal{E}$.}
    \label{fig:sigma_par}
\end{figure}

\textbf{Enhancement width $\sigma$:} \cref{fig:sigma_par} (a) and (b) show that for both Neumann and Dirichlet boundaries, $t_m$ is reached more quickly for narrower peaks; i.e. for smaller values of $\sigma$. This difference is only slight. There are two aspects at work here; a wider $\sigma$ will have access to larger diffusion rates at high $L$, but the gradient $f_L$ will be less steep. We can examine these components using $\mathcal{N}$ and $\mathcal{E}$ to determine which is more significant.

An enhancement across more $L$ has somewhat more mass and appears to los\old{s}\new{e} mass more quickly for both Neumann and Dirichlet outer boundary conditions. Despite the fact that experiments with a larger $\sigma$ also start with higher $\mathcal{N}$, at the end of the week, 98\% and 75\% of the mass remains from the initial population (for Neumann and  Dirichlet experiments respectively), compared to 99\% and 76\% remaining from our default experiments. Both experiments with a wider enhancement lose more from the inner boundary (\cref{fig:sigma_par}(d)), again showing that the initial condition controls more loss from the inner boundary than changes in the outer boundary condition. Loss from the outer boundary is slightly more with a larger $\sigma$, but is roughly comparable.

 Total $\mathcal{E}$ is higher for a larger $\sigma$, which makes sense as the experiment has more mass and hence larger $\int f^2 dL$; the distribution $f$ is further from a steady state. All $\mathcal{E}_t^{(1,3)}$ with $\sigma=0.5$ are slightly larger, but overall very similar. Unsurprisingly, with similar $\mathcal{E}_t$ but a larger starting $\mathcal{E}$, experiments starting with a larger $\sigma$ end up retaining proportionally more of the initial energy (37\% and 58\% for Dirichlet and Neumann respectively, rather than 30\% and 47\% of the initial energy when using default $\sigma$). Overall, with a wider $\sigma$, the Dirichlet experiment loses more $\mathcal{E}$, even though $t_m$ is slower. The results from the investigation of $\sigma$ indicate that the trade-off between gradient in the PSD and the $L$-dependence of $D_{LL}$ is subtle and nuanced (a wider enhancement has a less steep gradient but samples higher $D_{LL}$).

\subsubsection{Gradients versus the $L$-dependence of $D_{LL}$}\label{sec:gradients}
\begin{figure}
    \centering
    \includegraphics[width=\linewidth]{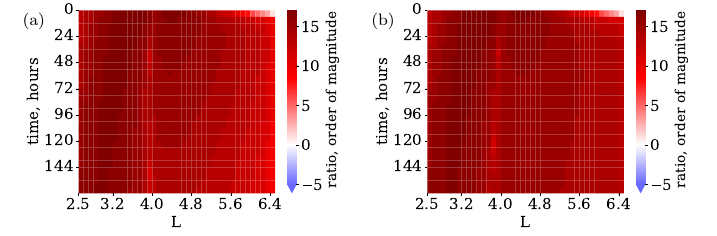}
    \caption{What contributes most to the reconfiguration term $\mathcal{E}_t^{(3)}$ (which dominates the evolution of the system)? We compare the two terms $(f_L)^2$ and $D_0 L^{n-2}$, for $n=6$. (a) is Neumann, (b) is Dirichlet\new{. We find an overwhelming dominance of the gradient term over diffusion coefficient, regardless of outer boundary condition.}}
    \label{fig:E3_oom}
\end{figure}

The $\sigma$ results suggest we should compare the role of the spatial (i.e., $L$) dependence and the gradients in the distribution function on the overall amount of diffusion. We consider their role in reconfiguration term $\mathcal{E}^{(3)}_t$, since this generally dominates $\mathcal{E}_t$. With a constant diffusion coefficient, only the gradient term would contribute to the $\mathcal{E}^{(3)}$. With an $L$-dependent $D_{LL}= D_0 L^n$, both $(f_L)^2$ and $D_0 L^{n-2}$ will contribute to $\mathcal{E}^{(3)}$. We simply compare the order of magnitude of these components via \old{a}\new{the} ratio\new{' of $(f_L)^2$  to $D_0 L^{n-2}$ for $n=6$}, shown in \cref{fig:E3_oom}. Overwhelmingly, it is the gradient component $(f_L)^2$ that dominates. This is unsurprising once one considers that $D_0 \sim 10^{-10}$ s$^{-1}$.

Throughout this analysis we have found that one must consider the whole domain; for example, the amount of diffusion is not limited by the smallest $D_{LL}$ but also the shape of the distribution, loss, the choice of domain etc. This is because $\mathcal{E}^{(3)}$ is the dominant component of $\mathcal{E}_t$, which determines the PSD evolution. And $\mathcal{E}^{(3)}$ is an integral over the entire simulation domain. 

%
%
%
%
\cref{fig:E3_oom} \old{shows the order of magnitude of the ratio of $(f_L)^2$  to $D_0 L^{n-2}$ for $n=6$. Over the whole domain, the gradients clearly have more impact than the L-dependence of the diffusion coefficient.}\new{ indicates that the gradients have more impact than the L-dependence of the diffusion coefficient.} However, with a longer \new{spatial ($L$) }domain, this will begin to change, particularly for a Neumann (zero flux) outer boundary, where there are fewer gradients. Using an idealised $D_{LL}$, gradients almost always dominate over the effect of $D_{LL}$ increasing with $L$. This will by why wider enhancements (larger $\sigma$) take longer to reach monotonicity when using our operational (Ozeke) $D_{LL}$: there are consequently shallower gradients [S1e]. 

We find that the PSD gradient $f_L$ contributes more to the evolution of the system than the diffusion coefficient $D_{LL} = D_0 L^6$\new{. Note that this idealised scaling was chosen to be comparable with the Ozeke diffusion coefficient at a given $L$ and Kp; there are many other models, often more sophisticated, yet all have a significant $L$-dependence. This is discussed further in \cref{sec:gradients_discussion}} [A2].

\subsubsection{Outer edge of domain, $L_{outer}$}

In this experiment we varied the domain for the simulation to see what difference it made. A Dirichlet (fixed value) condition is used in the majority of operational radiation belt models, to reflect observations\new{ e.g. \cite{SubbotinShprits2009,Glauert2018,Lee2024}}. The simulation domain is curtailed to the location of the spacecraft; different $L_{outer}$ values then correspond to using data from different spacecraft missions to set this outer boundary. Both types of outer boundary condition are investigated in this phase of experiments and the Dirichlet experiments retain the outer boundary value fixed in the initial condition.

\begin{figure}
    \centering
    \includegraphics[width=\linewidth]{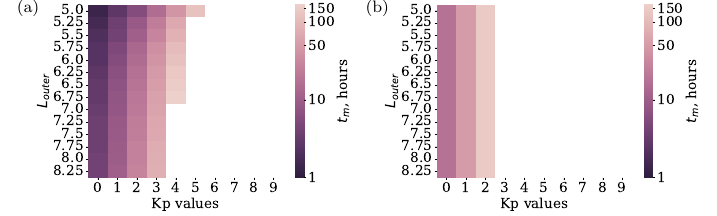}
    \caption{Time to monotonicity $t_m$ for $L_{outer}$ vs Kp over a week, for Neumann (left) and Dirichlet (right) outer boundary conditions. }
    \label{fig:Louter_tm}
\end{figure}

For a Neumann outer boundary condition, a more distant outer boundary (larger $L_{outer}$) took longer to reach monotonicity; i.e a smaller domain reached monotonicity more quickly (\cref{fig:Louter_tm}(a)). This suggests that the choice of outer boundary location changes the shape of the PSD distribution\new{, especially the height of the plateau}. For Dirichlet conditions, $t_m$ was independent of domain size.

To investigate this, we compare two Neumann runs where we vary the outer boundary location to be $L_{outer}=6.5$ and $7.5$. (Equivalent plots for Dirichlet can be found in the supplementary materials\new{, Figure S9)}. \cref{fig:N_cutthroughs} shows the PSD distribution at eight equally distant times throughout the week, with Kp=4 and using Ozeke $D_{LL}$. 
\begin{figure}
    \centering
    \includegraphics[width=\linewidth]{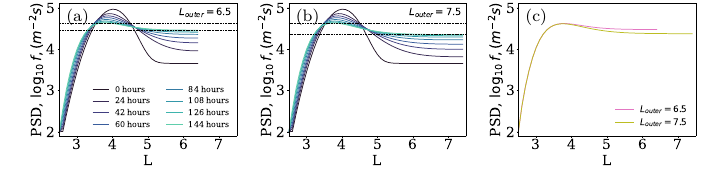}
    \caption{(a) Eight phase space density distribution snapshots from a week long simulation, with a Neumann (zero gradient) boundary at $L=6.5$. (b) Eight snapshots with the boundary at $L=7.5$.\new{ Dotted lines indicate the height of the peak and plateau by the end of the week.} (c) The final distribution for each of those.}
    \label{fig:N_cutthroughs}
\end{figure}
The \old{distance}\new{difference in PSD} between peak, and plateau edge (both indicated with dotted lines) by the end of the week is much larger for the run with a wider domain, in \cref{fig:N_cutthroughs}(b). The effect of this can be seen more clearly by considering the PSD at the final timestep, in \cref{fig:N_cutthroughs}(c). The run with $L_{outer}=6.5$ is more close to monotonic because the value of the outer boundary has raised higher. Despite the larger $D_{LL}$ at high $L$, more reconfiguration of the distribution is needed for a longer domain governed by the same equation - and as was seen in \cref{sec:gradients}, the gradients are still more important than the $D_{LL}$ up to $L=6.5$. Although the material in the peak being diffused outwards can be spread across more $L$ when there is a more distant $L
_{outer}$, and large $D_{LL}$ values at high $L$ encourage this, the material has to travel farther before the plateau rises and monotonicity is reached. The Neumann $t_m$ dependence on domain (i.e. on $L_{outer}$ arises because the shape of the distribution varies with changes in the simulation domain.

\begin{figure}
    \centering
    \includegraphics[width=\linewidth]{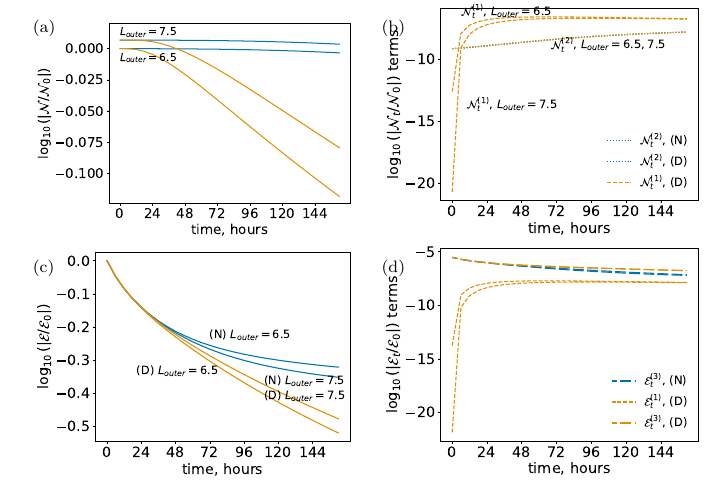}
    \caption{The changing mass-like and energy-like quantities $\mathcal{N}$  and $\mathcal{E}$ when the outer boundary location and condition are varied. The default $L_{outer}=6.5$ is compared to $L_{outer}=7.5$. The top row shows the total $\mathcal{N}$ and the components of the time derivative $\mathcal{N}_t$. The bottom row shows the same for the $L_2$ norm $\mathcal{E}$.}
    \label{fig:Louter_NE}
\end{figure}

Analysis of $\mathcal{N}$ in \cref{fig:Louter_NE} (b) indicates that neither the outer boundary location or condition affects loss from the inner boundary. For Dirichlet experiments with varying $L_{outer}$, mass loss from the outer boundary is of comparable order within a few hours, regardless of where that outer boundary is. This corroborates findings in \cref{sec:gradients} that the gradients in the PSD distribution dominate diffusion over the entire domain, rather than higher diffusion coefficients located in one region. While the extra mass at time $t=0$ was obvious for a longer domain, the change in initial $\mathcal{E}$ is negligible, as can be seen in \cref{fig:Louter_NE}(c). However, the evolution of the $L_2$ norm is nuanced; despite having $\mathcal{E}_t$ terms of similar order (\cref{fig:Louter_NE}(d)), with a longer domain, the Neumann experiments reaches a lower level of $\mathcal{E}$, while the Dirichlet experiment has a greater value of $\mathcal{E}$. 

The mechanism behind these results are, unsurprisingly, the rising plateau for Neumann and the outer boundary flux for Dirichlet experiments. In order for Neumann experiments to reach a configuration of lower $\mathcal{E}$, the high-$L$ plateau rises. With a longer domain, this means that there are more particles at high $L$, hence $\mathcal{E}$ can be lower than with the same number of particles at a lower $L$. Additionally, $\mathcal{E}^{(3)}$ is slightly larger for $L_{outer}=7.5$ than $6.5$, which can be attributed to outward radial diffusion due to the longer domain and higher $D_{LL}$ at $L=7.5$. For the Dirichlet case, the reconfiguration energy change is almost exactly the same. However,  more is lost to the outer boundary for $L_{outer}=7.5 $ than $6.5$. Even though they lose roughly the same each hour after 70 hours, this is enough to make $\mathcal{E}$ slightly lower for $L_{outer}=6.5$

Using $\mathcal{N,E}$, we find that the choice of outer boundary location changes the shape of the PSD distribution for Neumann; in exactly the same manner, $t_m$ varies depending on the outer boundary location. For different outer boundary conditions, a different outer boundary location could result in heading faster or slower to a state where the dynamics are minimally changing (i.e. to minimum $\mathcal{E}$). In \cref{sec:Discussion} we discuss outer boundary choices, including the applicability of using a Dirichlet outer boundary that is fixed, but not using observations [P1c].

\subsection{Results Part 2: Including loss rate}
Loss from pitch angle scattering is significant and should be included to see how it relates to the timescale of radial diffusion. We include this loss by modelling the electron lifetime.

\subsubsection{The difference between the two outer boundary conditions, with loss}\label{sec:NvsD_loss}
In general, with loss it is no longer always true that more Neumann experiments reach monotonicity; nevertheless, they still have shorter $t_m$ than Dirichlet experiments. All the $t_m$ plots can be found in supplementary materials; again we select the results that inform us about the overall pattern.

\begin{figure}[ht]
    \centering
    \includegraphics[width=\linewidth]{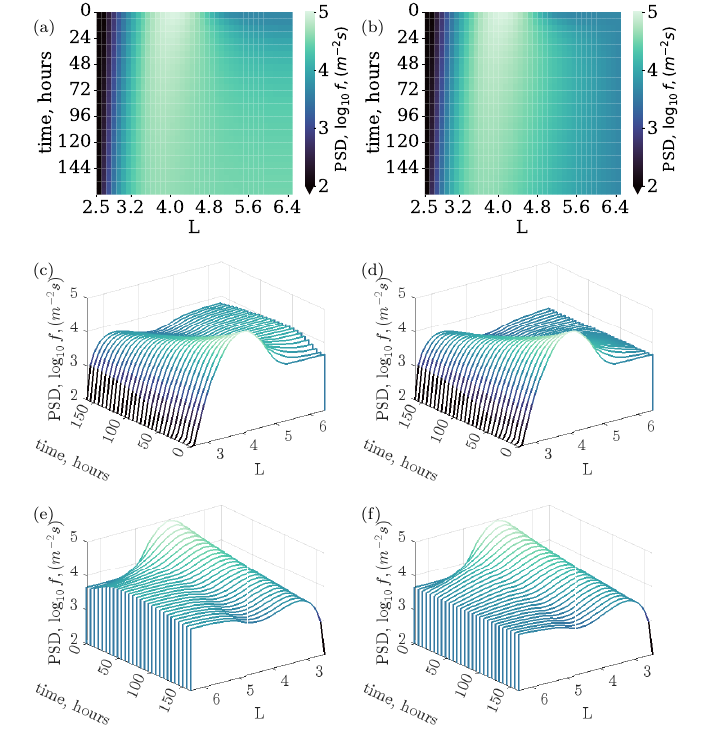}
    \caption{\new{Phase space density (PSD) over a week, with Neumann (constant
gradient, left) and Dirichlet (constant value, right) outer boundaries, where loss from pitch angle scattering has been included.}\old{Left: Neumann. Right: Dirichlet.} Row 1: Heatmaps of \old{log$_{10}$PSD}\new{PSD}. Row 2: waterfall plots (same as \cref{fig:PSDtime}). Row 3: waterfall plots from the back.}
    \label{fig:PSDtime_loss}
\end{figure}

\cref{fig:PSDtime_loss} show experiments with pitch angle loss for the default initial condition. Neumann and Dirichlet runs look very similar, suggesting that pitch angle loss may control more of the dynamics than the outer boundary condition. The analytic quantities $\mathcal{N,E}$ for these runs confirms this; 
\begin{figure}
    \centering
    \includegraphics[width=\linewidth]{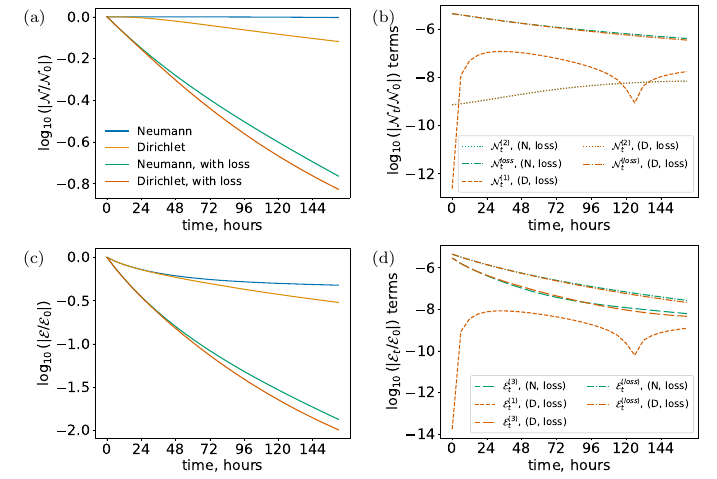}
    \caption{Same as \cref{fig:NvsD_EN} showing the difference between Neumann and Dirichlet experiments but with additional experiments containing loss from pitch angle scattering. The second column (b) and (d) showing mass and energy moment change terms only the terms for the loss experiments for readibility; full versions can be found in the supplementary materials\new{\ Figure S7 ($\mathcal{N}_t$) and Figure S8 ($\mathcal{E}_t$)}.}
    \label{fig:NvsD_EN_loss}
\end{figure}
loss from the pitch angle scattering approximation dominates over loss from the outer or inner boundary \cref{fig:NvsD_EN_loss}(b), and the evolution of $\mathcal{E}$ with loss looks similar regardless of outer boundary condition, unlike the runs without loss (\cref{fig:NvsD_EN_loss}(a)). The experiments including pitch angle loss are heading much more quickly towards a steady state, and after around a hundred hours the experiments are not changing much in $\mathcal{E}$, as $\mathcal{E}_t \sim 0$.

$t_m$ is more complex to analyse. Over the span of the week, enough mass is lost that the Dirichlet run begins to gain mass. Just as without loss, particles entering the domain are contributing to an increase in $\mathcal{E}$ which results in a final $\mathcal{E}$  that is higher for Dirichlet than Neumann. As a result, the Neumann case is closer to a steady state. Loss from pitch angle scattering effectively mimics the reduction in PSD that would occur over a very long timescale of radial diffusion, as it is strongest near the edge of the plasmapause (which is located around the bulk of the enhancement). With a constant inflow of particles, Dirichlet runs have a higher $\mathcal{E}$ but do not come closer to monotonicity. Electron lifetime loss is creating a new local minimum in the PSD, and so the distribution is not monotonic. To demonstrate what is happening in \cref{fig:NvsD_EN_loss} for both Neumann and Dirichlet runs, the diagram in \cref{fig:diagram4}(a) shows an example phase space density distribution with this additional minimum. This physical profile is corroborated by the fact that $\mathcal{E}_t^{loss}> \mathcal{E}_t^{(3)}$ and therefore loss dominates over reconfiguration in the syste\old{l}\new{m} evolution. We will expand on the consequences of this below. Finally, there is a difference in inner boundary flux; all experiments including lifetime loss have the same, lower flux [S1b;S4].
\begin{notestocoauthors}
    {\color{purple}cut: Loss is a larger factor than reconfiguration; furthermore, by the end of teh week, the $\mathcal{E}_t^{(3)}$ term is smaller than it was without loss. Probably because the loss is strongest where there happen to be the steepest gradients, so when these gradients are eroded, $\mathcal{E}_t$ gets boring.}
\end{notestocoauthors}

\subsubsection{When $t_m$ can never be reached: how $\mathcal{L}$ affects monotonicity}\label{sec:never_monotonic}
Although experiments with loss $\mathcal{L}$ that reach monotonicity do so quicker than they did without $\mathcal{L}$, for all parameters there are several initial values that reached monotonicity without loss but no longer do once loss is included, usually at lower Kp (i.e. weaker radial diffusion relative to the same loss). The reverse is rarely true; only for narrow $\sigma$ \new{(e.g. 0.2, 0.25) }and $L_{outer}=5.0$ is a $t_m$ found with $\mathcal{L}$ where none was found before. In this section we explain the physical mechanism behind this general pattern; again, all individual $t_m$ plots can be found in the supplementary materials, Figure S5.

\begin{figure}[ht]
    \centering
    \includegraphics[width=\linewidth]{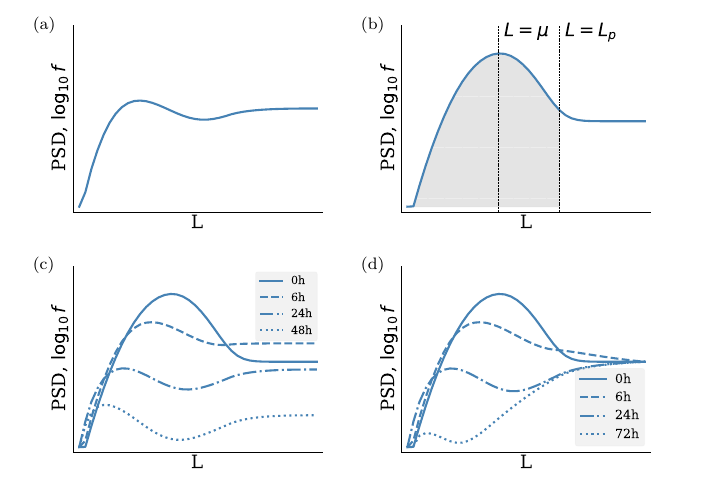}
    \caption{(a) shows a PSD with an enhancement, where pitch angle scattering at higher $L$ than the enhancement has resulted in a new minima. (b) shows the case where this occurs; when the plasmapause is at higher $L$ than the enhancement ($L_p >\mu$) then the loss region (shaded) includes the area to the right of the enhancement, which can prevent reaching monotonicity. The second row shows how the PSD distribution evolves when loss dominates over the diffusion; the distribution may never reach monotonicity but will look different for Neumann (left) and Dirichlet (right) outer boundary conditions. The second row is from experiments with the default initial condition, $Kp=4$ and $L_p=5$.}
    \label{fig:diagram4}
\end{figure}

The existence of experimental setups that will never reach monotonicity is most clearly demonstrated by following a case where $L_p > \mu$ (\cref{fig:diagram4})(b). There will be loss everywhere left of $L_p$. In the region $\mu<L<L_p$, if loss $\mathcal{L}$ is strong enough it can work against monotonicity by creating a minimum. The second row of \cref{fig:diagram4} demonstrates how this case can evolve for Neumann and Dirichlet outer boundary conditions, using default initial conditions, the same loss with the default plasmapause at $L_p=5$ and $Kp=4$ in the diffusion coefficients. The Neumann and Dirichlet (\cref{fig:diagram4}(c) and (d) respectively) experiments show that even when the distribution is much reduced (over 24 and 72 hours respectively), it is not monotonic. (Note that where the loss does not dominate over radial diffusion, then instead the effect of the loss is for the overall distribution to reduce more quickly, but still maintain the characteristic diffusion distribution (e.g. the intermediate distribution in \cref{fig:diagram1and2}(b)))[S3a;S3b] 

\subsubsection{Loss affects the evolution of diffusion more than most properties of the initial condition}
In general, initial conditions impact the diffusion in a similar manner to without loss, for example $t_m$ varies significantly with $\mu,B$. In this section, variation of each parameter is compared to the case without loss.  Again, we select the most significant results here, while all the figures can be found in the supplementary materials\new{\ Figures S5-8}. \old{Note that each $\mathcal{N,E}$ is normalised using the initial $\mathcal{N}_0,\mathcal{E}_0$ for the high-parameter experiment without loss, whereas in 
Section V A they were normalised against the initial values  $\mathcal{N}_0,\mathcal{E}_0$ for the phase space density using all default parameter values.}\new{In }\cref{sec:resultspart1}\new{\ each $\mathcal{N,E}$ was normalised using initial values $\mathcal{N}_0,\mathcal{E}_0$ for the phase space density using all default parameter values for the initial condition. To compare the effect of loss, we instead choose normalisation values $\mathcal{N}_0,\mathcal{E}_0$ from the initial phase space density of the higher parameter value in each experiment pair, e.g. we normalise using the initial mass and energy density with $B=5$ rather than $B=2$, $\mu=5$ rather than $\mu=4$, etc.} This normalisation was chosen to ensure that the effect of each parameter is extracted, rather than \new{repeating experiments comparing the default initial condition with and without loss,}\old{including more of the generic evolution with loss} which was explored in \cref{sec:NvsD_loss} and \cref{sec:never_monotonic} [P1b;S3a;S4].

\begin{figure}
    \centering
    \includegraphics[width=\linewidth]{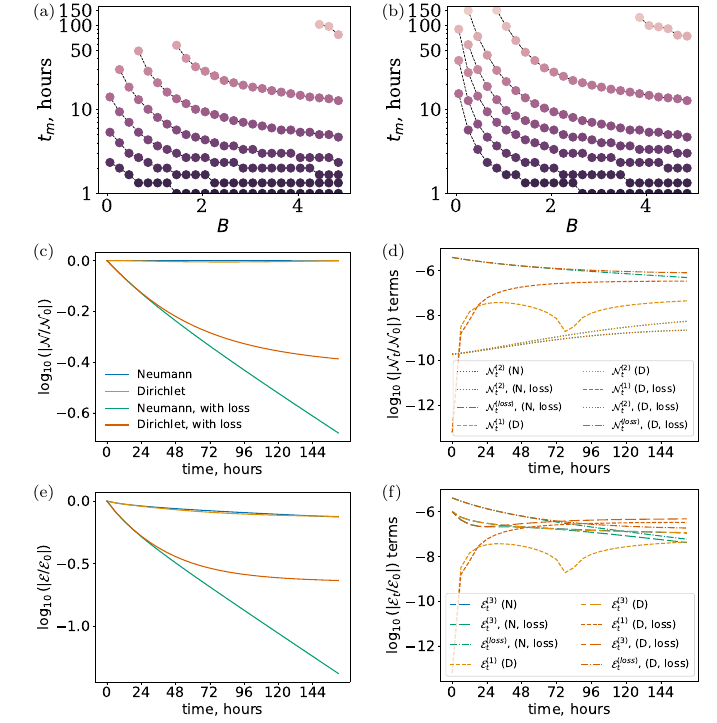}
    \caption{Same as \cref{fig:B_par} but compares the higher step size $B=5$ experiments to their equivalents with loss from pitch angle scattering. Note that the outer boundary terms $\mathcal{N}_t^{(1)},\mathcal{E}_t^{(1)}$ for the loss experiments represent gains as material flows into the domain.}
    \label{fig:B_par_loss}
\end{figure}

\textbf{Step $B$:} Just as without loss, a higher step size $B$ results in a shorter $t_m$ as the distribution is already closer to monotonic. However, this effect is much smaller. As can be seen in \cref{fig:B_par_loss}(a) and (b), the $t_m$  relationship to changing $B$ is still exponential, but stops changing once $B\leq2$. In fact, a step this large quickly results in gains from the outer boundary; in \cref{fig:B_par_loss}(d) and (f), $\mathcal{N}^{(1)}$ and $\mathcal{E}^{(1)}$  for experiments with loss are always positive. The Dirichlet experiment also finds that at around 72 hours, the reconfiguration term and outer boundary terms dominate over the loss term for the $L_2$ norm, $\mathcal{E}^{(1)},\mathcal{E}^{(3)} > \mathcal{E}^{(loss)}$. As a result, the high-$B$ Dirichlet experiment reaches a state where the constant churn of material being brought into the domain and diffused inwards dominates over the loss from pitch angle scattering. The Neumann experiment obviously does not experience this. Because the Dirichlet simulation has quickly reached a point where the dynamics are no longer changing (due to the large fixed value of $f_{OB}$) the Neumann and Dirichlet experiments diverge in $\mathcal{E}$ even though in general, the outer boundary condition has less effect than loss.


\textbf{Enhancement location $\mu$:} \old{As without loss, a higher $ \mu$ means that $t_m$ is reached more quickly. Fewer experiments reach monotonicity; those that do reach it more quickly. For the first few hours, the loss experiments actually find that $\mathcal{E}_t$ is dominated by reconfiguration - this will be because the enhancement is at higher $L$ so more diffusion is possible. Then loss becomes dominant - for Neumann with loss, the reconfiguration and loss contributions to $\mathcal{E}_t$ become comparable after around 70 hours. ``Reconfiguration'' $\mathcal{E}_t^{(3)}$ is not as strong as without loss.}\new{As without loss, a higher $ \mu$ means that $t_m$ is reached more quickly.\footnote{We also note the minor effect from Figure S8 (f) that for the first few hours, the loss experiments actually find that $\mathcal{E}_t$ is dominated by reconfiguration $\mathcal{E}_t^{(3)}$ - this will be because the enhancement is at higher $L$ so more diffusion is possible. Then loss becomes dominant - for Neumann with loss, the reconfiguration and loss contributions to $\mathcal{E}_t$ become comparable after around 70 hours. ``Reconfiguration'' $\mathcal{E}_t^{(3)}$ is not as strong as without loss.}}


\textbf{Amplitude $A$:} There is no change with\new{\ overall} amplitude \new{parameter $A$\ }- the same as without loss. We see \new{the\ }same \old{overall}\new{general} patterns in $\mathcal{N},\mathcal{E}$ as discussed in previous section.

\textbf{Enhancement width $\sigma$:} Again, a narrower width (i.e. lower $\sigma$) reaches $t_m$ quicker, the same as without loss. From $\mathcal{N,E}$ analysis (shown in supplementary materials\new{, Figures S7 and S8, (g) and (h)}) we find the same overall results as discussed in \cref{sec:NvsD_loss}. With loss, the Neumann and Dirichlet runs are more similar to each other than to the runs without loss. With loss, less is lost through the inner boundary.

\subsubsection{Outer edge of domain, $L_{outer}$}
\begin{figure}
    \centering
    \includegraphics[width=\linewidth]{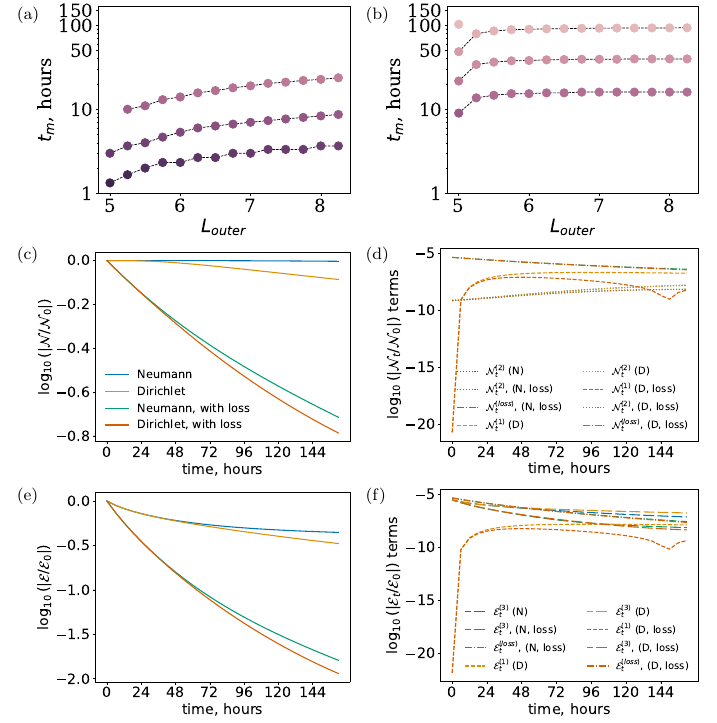}
    \caption{Same as \cref{fig:B_par_loss} but investigating the interaction between $L_{outer}$ and loss\new{, for $L_{outer}=7.5$}.}
    \label{fig:Louter_par_loss}
\end{figure}

Without loss, we found that $t_m$ varied when we changed the location of a Neumann outer boundary condition, but not a Dirichlet condition. With loss, we find that $t_m$ dependence on $L_{outer}$ is very small for both a Neumann boundary (\cref{fig:Louter_par_loss}(a)) and a Dirichlet boundary, particularly when the domain boundary and the plasmapause are close (\cref{fig:Louter_par_loss}(b))[P1c].

\subsubsection{Plasmapause location $L_p$}

\begin{figure}
    \centering
    \includegraphics[width=\linewidth]{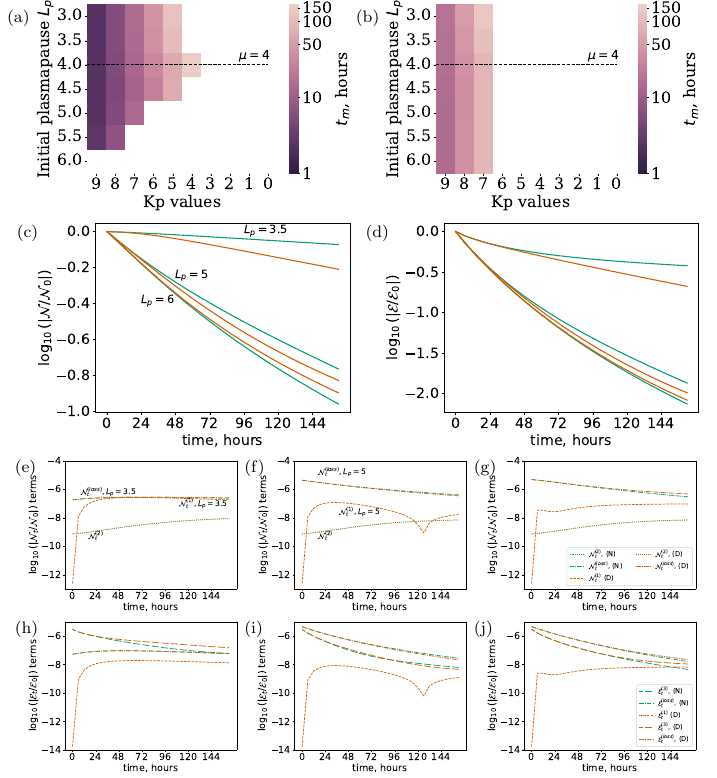}
    \caption{First row: time to monotonicity for Neumann (left) and Dirichlet (right) outer boundary conditions with varying Kp and plasmapause location $L_p$.\new{\ A dotted line indicates the default peak position $\mu=4$.} The second row shows the combined $\mathcal{N}$ (c) and $\mathcal{E}$ (d) results for three plasmapause locations, $L_p=3.5,5,6$, across a week-long simulation.\new{\ Green lines indicate Neumann (`N') outer boundary condition experiments with loss, while the dark orange lines indicate Dirichlet (`D') experiments.} The third row shows the $\mathcal{N}_t$ terms for each of these respectively, and the fourth row shows the $\mathcal{E}_t$ terms. Note that outer boundary contributions for $L_p=6$ (i.e. $\mathcal{N}^{(1)},\mathcal{E}^{(1)}$) are positive.}
    \label{fig:Lp_par}
\end{figure}

When including loss, the extent of the lossy region is a new parameter to consider. The outer limit of this region is the plasmapause, $L_p$.\new{\ By default our simulations have $L_p=5$, which is higher in $L$ than the default peak location, $\mu=4$.\ } \cref{fig:Lp_par}(a) and (b) show time to monotonicity across a variety of plasmapause locations. The effect for a Dirichlet outer boundary condition is small, with an effect for the lowest plasmapause values (for which $t_m$ is sooner) which quickly drops off to reach a value of $t_m$ that no longer changes with $L_p$. The effect of plasma\new{p}ause location with Neumann outer boundary condition is more complex; overall, a plasmapause closer to the Earth reaches monotonicity sooner. However, there is a cut-off in $L$ after which monotonicity is not reached at all\new{; our experiments place this around $L_p=6$. This is unlikely to be a ``hard'' cut-off but instead due to interactions with $\mu$ and $L_{outer}$. These relationships are explored in the following paragraphs}.

In \cref{sec:never_monotonic} we noted that $L_p > \mu$ could result in the PSD being unable to reach a monotonic state. This relationship is explored by using three plasmapause locations in the $\mathcal{N,E}$ analysis: $L_p= 3.5, 5,6.$ These results are shown in the final three rows of \cref{fig:Lp_par}.

As the default enhancement location is $\mu=4$, a plasmapause at $L=3.5$ is at a lower $L$ than the enhancement. \cref{fig:Lp_par}(e) indicates that material lost from pitch angle scattering is quickly similar to loss from the outer boundary, and (h) demonstrates that $\mathcal{E}_t$ is dominated by the reconfiguration term $\mathcal{E}^{(3)}$ (although this becomes comparable to $\mathcal{E}^{(loss)}$ for the Neumann experiment by the end of the week). With less overall loss, the Neumann and Dirichlet experiments still have distinct values of $\mathcal{N,E}$.

For a plasmapause at $L=5$ or $6$, much more material is lost, as expected since the proportion of particles lost increases with $L$. The norm for $L_p=5,6$ is quickly very low (\cref{fig:Lp_par}(d)). For $L_p=6$, the Neumann experiment reaches a lower energy state, whilst for $L_p=5$ the Dirichlet experiment has a lower norm. This is due to more material coming in the outer boundary with a higher $L_p$. 

\begin{figure}
    \centering
    \includegraphics[width=\linewidth]{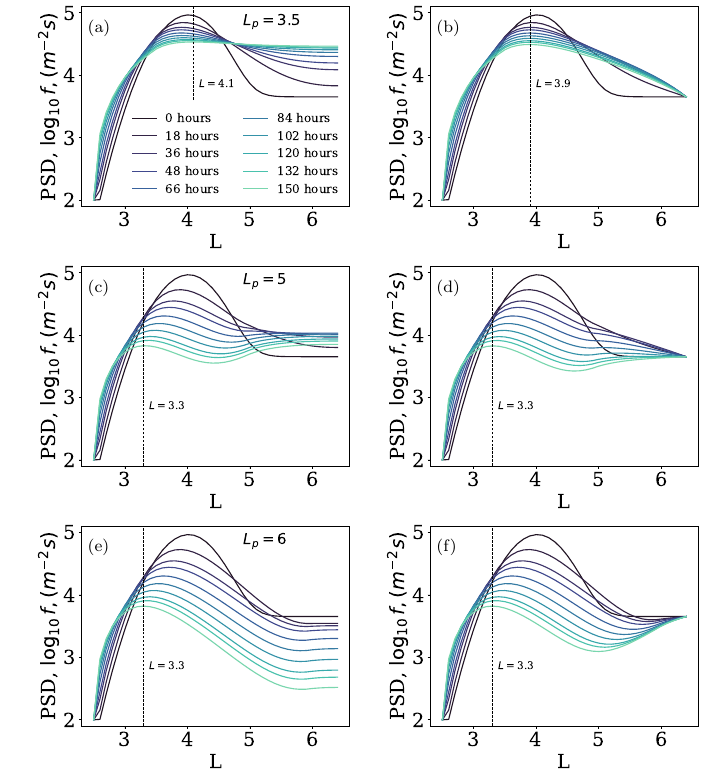}
    \caption{Intermediate phase space density distributions from week long experiments, with a plasmapause $L_p$ at $3.5$ (row 1), $L_p=5$ (second row) and $L_p=6$ (third row). Runs on the left and right have Neumann and Dirichlet outer boundary conditions respectively.\new{\ Dotted lines indicate the location of the peak of the enhancement (the maximum PSD value across lower $L$) by the end of the experiment.)}}
    \label{fig:Lp_cuts}
\end{figure}

\new{The relative location of peak and plasmapause are the determiners of whether a local minimum is at all possible, while details of $\mu$, $\sigma$ and $L_p$ combine to see whether it occurs in a given simulation. For example , despite the entire enhancement lying inside the plasmapause for $L_p=6$, we see that the local minimum in \cref{fig:Lp_cuts} (e) is very small, even though a local minimum very clearly occurs when the peak is just inside the plasmapause (e.g. $\mu=4, L_p=5$, \cref{fig:Lp_cuts} Row 2). In this case, the loss from pitch angle scattering results in the peak rapidly moving inwards, and a local PSD minimum arising between $L=4$ and $L=5$ (\cref{fig:Lp_cuts} (c) and (d)). Peak width may change the rate at which a local minimum arises (e.g. a wider enhancement could mean a higher PSD to lose before reaching a local minimum, while a narrower enhancement could mean that diffusion occurs more quickly, offsetting the pitch angle loss) but it is the relative location of $\mu$ and $L_p$ that determine whether this is possible.}

For a higher plasmapause location, $\mathcal{E}^{(loss)}$ becomes increasingly larger. When $\mathcal{E}^{(loss)}>\mathcal{E}^{(3)}$, diffusion can not prevent the formation of the extra minimum demonstrated in \cref{sec:never_monotonic}. Indeed, this is the relationship observed when plotting several intermediate time instances of the experiments with $L_p=3.5,5,6$, shown in \cref{fig:Lp_cuts}. For $L_p<\mu$, this minimum does not form (first row of \cref{fig:Lp_cuts}) while the minimum is larger as $L_p$ increases (\cref{fig:Lp_cuts} second and third rows).

Note that since $\mathcal{E}^{(3)_t}$ depends on the gradient of the PSD, the growth of this minima (i.e. when $\mathcal{E}^{(loss)}_t> \mathcal{E}_t^{(3)}$) is dependent on the initial conditions. Loss may or may not dominate the evolution of the system over reconfiguration. Overall, our results emphasise that the plasmapause location relative to the enhancement is important, and that a more distant plasmapause has so much more loss that this can totally change the dynamics [P1b; S1b; S2; S4].

\subsection{Summary of Results: The role of the Initial Condition}


Monotonicity was easier to obtain when there was already a significant background PSD, corresponding to a large high-$L$ source (i.e. high $B$). \old{Therefore, $t_m$ corresponds to the timescale of an enhancement relative to background field.}\new{Therefore, $t_m$ corresponds to the timescale of radially diffusing a local enhancement, with respect to the background PSD (the background particle population).} If the enhancement location $\mu$ occurs at high $L$, it doesn't last as long before the distribution becomes monotonic\new{, although including loss from pitch angle scattering means that the relative location of the peak and plasmapause strongly interact to affect $t_m$}. Furthermore, a narrower enhancement will be reduced more quickly than a wider one - and using $\mathcal{E}_t$, we attributed this to the gradients of the PSD, which will be discussed in \cref{sec:gradients_discussion}. Less intuitively, we found that the time for our enhancement to ``fade" into the background varies significantly with the outer boundary location if we used a Neumann boundary. This is a key result, and we explore the consequences of this and future avenues in \cref{sec:Louter_discussion}. Finally, we found that loss from pitch angle scattering has a strong effect on the final shape of the distribution (and therefore the timescale for radial diffusion). Indeed, some numerical experiments never reach monotonicity, casting some questions about future appropriateness of $t_m$, which we discuss in \cref{sec:eval_tmNE} [P1b;S1b;S2;S3a;A2].

Using the evolution of $\mathcal{N}$ we could work out what processes were going on, and from $\mathcal{E}$ we could work out how the system was evolving to reach a maximally-diffused state. We found that although theoretically an experiment with a Neumann OBC can reach a state with a lower $L_2$ norm (i.e. zero everywhere), the majority of the time Dirichlet experiments were reaching a state of lower $\mathcal{E}$, because mass could be lost from both boundaries. The Neumann experiments were reaching a state where $\mathcal{E}_t\sim 0$; where the dynamics were changing very little once the plateau had risen up, whilst the Dirichlet experiments were still diffusing material from the enhancement by the end of the week [S1b;A1]. 

Other specific results from the use of $\mathcal{N,E}$ include the importance of gradients in the distribution; using $\mathcal{E}_t^{(3)}$ we can estimate the comparative effect of the $L-$dependence of $D_{LL}$ and the gradients in the distributi\old{u}on on the evolution of $f$. We found that gradients are more significant, a result worthy of its own discussion section \cref{sec:gradients_discussion}. We can also see in \cref{fig:Louter_NE}(c) that although using a different domain (i.e. a different $L_{outer}$) does not significantly impact the starting value of $\mathcal{E}$, it \textbf{does} affect the rate at which the system moves towards a steady state (more in \cref{sec:Louter_discussion}, \cref{sec:timescale_discussion}). Finally, we can quantify that loss from pitch-angle scattering has a stronger effect on the evolution of $f$ than radial diffusion does [S4;A2]. 

\section{Discussion: Implications for Future Work}
\label{sec:Discussion} 


Understanding of the research goals, and the context of our results, has developed throughout the research lifetime of this work. The initial questions shaped the methodology and investigation of the results shaped the narrative of the analysis. Here we put the results into context with existing literature and with future modelling choices. To aid navigation, relevant paragraphs are labelled with our initial research questions in \cref{sec:goals}. Each research goal may be addressed in multiple paragraphs.

\subsection{Evaluation of $t_m$ and $\mathcal{N,E}$ as analysis tools}\label{sec:eval_tmNE}
Error metrics such as the log-accuracy ratio \cite{Morley2018} are often the first tools considered for comparing distributions; this would be a suitable method to compare the deviation between two \new{phase space density (PSD)\ }distributions\new{\ $f$}, such as between observations and models (although weighting by the Jacobian $1/L^2$ may be necessary, as it has been here). When no ``truth'' is available to compare against, error metrics become an unsuitable tool as it would require a threshold (e.g. when two distributions are ``close enough", or when radial diffusion is ``done") which would be difficult to motivate objectively. In this section we discuss the performance of tools \new{$t_m$ (time for PSD reach a monotonic state), $\mathcal{N}$ (mass density, \cref{eq:N}) and $\mathcal{E}$ (\cref{eq:E}, energy density, or ``distance from zero state'')}\old{$t_m,\mathcal{N,E}$} when comparing distributions and how the distribution morphology became so integral to our analysis. 

$t_m$ and $\mathcal{N,E}$ are found to be complementary measures of the shape and evolution of the distribution. They are related, as the state of monotonicity and the state of lowest possible $\mathcal{E}$ both have PSDs predominantly weighted towards the outer edge of the domain. $t_m$ is a state where the only existing gradients are to the left hand side of the enhancement. Indeed, $\mathcal{E}$ particularly penalises high $f$ at low $L$ (e.g. $f=1$ at at $L=4$ results in a higher $\mathcal{E}$ (1/16) than at $L=5$ (1/25)). (Note, however, that a monotonic distribution can be gaining particles from the outer boundary and therefore\new{\ be increasing} in $\mathcal{E}$). Despite the similarity between low-$\mathcal{E}$ and monotonic states, evolution is determined by $\mathcal{E}_t$; the steady state reached may not be the lowest $\mathcal{E}$ configuration possible. $\mathcal{E}_t$ is in turn usually dominated by reconfiguration term $\mathcal{E}^{(3)}_t$, which is an integral across $L$ that is also dependent on $L$. This term tells us that gradients in $f$ (and their location in $L$) are the strongest factors dominating $\mathcal{E}^{(3)}_t$. As a result, the distribution will change more rapidly towards the low-$\mathcal{E}$, monotonic-like state when there are steeper gradients, and when those gradients are situated at low $L$.

The dominance of $\mathcal{E}^{(3)}$ consequently suggests that quantities capturing the shape of the distribution and the location of features in it are going to be the most useful tools when analysing the output - as we have found. We conclude that measures using the distribution morphology across the entire domain are necessary to capture the system evolution. 
Indeed, the outer boundary condition affects the monotonicity of the final state. The shape of the distribution and the ongoing evolution change with the location and condition of the outer boundary. The act of ``reducing gradients” means that in general, Neumann experiments reach monotonicity quickly, whilst the steep gradients remaining in the Dirichlet experiments means that they are actually at lower energy states and still changing more rapidly (losing more $L_2$ norm and mass) towards a steady state.

We suggest that we finally settled on this pair of tools because they are (a) both domain dependent, and (b) include information about the morphology of the PSD distribution. Both tell us about the entire system,  i.e. $\mathcal{E,N}$ tells us about the system across all $L$ at each timestep; $\mathcal{N}_t,\mathcal{E}_t$ inform us about the ongoing evolution of the system at each timestep, and $t_m$ is a measure of the long-term evolution. Initially, we searched for domain \emph{in}-dependent measures; but as the diffusion coefficient $D_{LL}$ is domain dependent, so should be our method of analysis. Both tools relate, predominantly, to gradients. The existence of steep gradients are the largest contributor to a rapidly changing $\mathcal{E}$; a distribution rapidly heading towards a steady state. A moderate limitation of $\mathcal{N,E}$ is that one can only use idealised diffusion coefficients rather than the empirical ones from \cite{Ozeke2014} used in calculating $t_m$. Unfortunately, a limitation of $t_m$ is that some parameter combinations never reach a monotonic state, i.e. once loss is included at higher $L$ values (see \cref{sec:never_monotonic} and \cref{fig:Lp_cuts}). The extra minimum in these cases suggest that we may need to characterise by convexity (i.e. number and location of extrema) rather than simply monotonicity and $t_m$. Nevertheless, we found $t_m$ to be an extremely useful measure to compare the different experiments, which could be related back to radiation belt processes, to a ``quiet” state and to quantifiable properties of the PSD $f$. Using both $t_m$ and $\mathcal{N,E}$ together gives us a nuanced description of timescale (\cref{sec:timescale_discussion}). These quantities allowed us a clear and efficient means of comparing experiments and understanding the processes behind the evolution of each [S3a;S3b]



\subsection{Timescale for radial diffusion}\label{sec:timescale_discussion}
\begin{notestocoauthors} 
{\color{purple}Dan - you said something about a characteristic scale - how can we get that if the diffusion coefficinet is not not constant?}
\end{notestocoauthors}

A radial diffusion ``timescale" is poorly defined when $D_{LL}$ depends on $L$; one cannot simply take $\frac{1}{D_{LL}}$ as the characteristic timescale. This is an open question both for idealised models and the real radiation belts, because of the number of underlying approximations. Potential measures of timescale include the autocorrelation, or dimensional analysis of self similar solutions (e.g. \cite{OsmaneLejosne2021}). Unfortunately, self similar timescales are difficult to find for the 1D radial diffusion equation \cref{eq:diffusion} with diffusion coefficients \cref{eq:OzekeDLL} or \cref{eq:LnDLL} because of the high order of $L$. Typically, one makes ``by eye" judgments of storm time radiation belt simulations. One could numerically solve for the equilibrium solution and then find the time taken to reach this (e.g. \cite{Ripoll2016}) but this will also depend on the metric or threshold one chooses to define as close enough to this equilibrium solution. Alternatively, suggestions were to take either the minimum or maximum $1/D_{LL}$ in the system, as this would at least give you a ``fastest possible'' timescale (\textit{various personal communications}). We have explored methods of quantifying timescale using our tools.

Using $t_m$, the concept of timescale reduces to ``how long until an enhancement is diffused away". The results in \cref{sec:resultspart1} thereby include how the initial  condition (e.g. size and location of enhancement) relate to timescale without loss. There is a large variability with Kp$, B, \mu$ etc. From $t_m$ at a Kp of 4-5 we conclude that the timescale is on the order of days, or tens of hours (for all the $t_m$ plots together, see Figure S1). Some combinations of initial conditions will keep \new{$t_m$}\old{this} close to one day, others closer to a week.\new{\ We show some example timescales using each of these measures in \cref{tab:timescales}. Based on these results, we conclude that time to monotonicity is a more representative measure of timescale, but has practical limitations based on the outer boundary and on the fact that not all distributions can reach a monotonic distribution.} This definition is not particularly suited once one considers loss from pitch angle scattering; although the timescale drops to hours or days, since monotonicity is no longer guaranteed it is a poor indicator of timescale [P2a,b,c].
\begin{table}
\caption{\label{tab:timescales} \new{\textbf{Example radial diffusion timescales.} These experiments used an initial phase space density with an enhancement at $\mu=5$, both Neumann (fixed gradient, `N') and Dirichlet (fixed value, `D') outer boundary conditions and two outer boundary locations $L_{outer}=6.5$ and $7.5$. Experiments were run with and without loss from pitch angle scattering. The timescales shown here are the `maximum' timescale using the Ozeke diffusion coefficient ($1/\min{D_{LL}}$), the `minimum' timescale ($1/\max{D_{LL}}$) and time to monotonicity $t_m$. Kp$=4$ was used throughout. Entries are blank if monotonicity was not achieved within a week. Most entries are rounded to the nearest hour.}}
\begin{ruledtabular}
\begin{tabular}{lcccc}
  & Min. $1/D_{LL}$ & Max. $1/D_{LL}$ & $t_m$ (N) & $t_m$ (D) \\ 
 \hline
 $L_{outer} =6.5$ & 9 h & 18668 h & 40 h & - \\ 
 $L_{outer} =7.5$ & 0.2 s & 18668 h & 75 h & - \\ 
 \hline
 $L_{outer} =6.5$ (loss) & 9 h & 18668 h & 24 h & 92 h \\ 
 $L_{outer} =7.5$ (loss) & 0.2 s & 18668 h & 39 h & 111 h \\ 
\end{tabular}
\end{ruledtabular}
\end{table}

$\mathcal{E}$ is even less suited to extracting a timescale as one would need to set a threshold. Nevertheless, we can draw some qualitative conclusions. Without loss, within a few days we see that the Neumann ensembles for all initial conditions have a relatively flat $\mathcal{E}$ - the dynamics are not changing. On the other hand, most Dirichlet experiments are still reducing in $\mathcal{E}$ at this point. Obviously this timescale is difficult to use as it has such a strong dependence on the outer boundary condition - which reflects the importance of the outer boundary condition. With loss, we see that initial conditions matter far less; with 72 hours most ensemble members have reached a very low energy, even though they are still losing mass; the timescale of the dynamic processes is only a few days before becoming ``quiet" [P2a]. 

Our measures of timescales tell us (a) when the enhancement is diffused to the background via $t_m$ (although needs adapting for the case with loss) and (b) whether the dynamics are still changing (i.e. whether there are rapidly changing gradients being diffused away) via $\mathcal{E,E}_t$. We find that pitch angle loss generally dominates over the radial diffusion timescale (and note that in practice, pitch angle scattering timescales also vary with $L$ and energy, \cite{Ripoll2016}).


As a result, from our work we can suggest only that the radial diffusion timescale is on a scale of hours to days, depending on the parameters one uses (e.g. Kp, $\mu$ etc). We note that using either the minimum or maximum value of $\frac{1}{D_{LL}}$ is not a good indicator of timescale, for two reasons: (a) our conclusion that one should consider the entire domain, and (b) our result that gradients in the PSD affect the evolution of the system more than the value of $D_{LL}$, at least for a reasonable activity magnetosphere ($Kp=4$) and up to $L=6.5$. The definition of timescale is also still poorly defined and should be specified for a given purpose in order for any quantitative conclusions to be reached, e.g. time for an enhancement to be diffused away, for the radiation belts to drop below a certain energy, or return to a specific state. One interesting potential timescale would be the time taken for loss or reconfiguration terms ($\mathcal{E}^{(loss)}$, $\mathcal{E}^{(3)}$) to dominate evolution (i.e. to be the dominant term of $\mathcal{E}_t$) [P2a;P2b;A3].



\subsection{Ensembles for radiation belt modelling}\label{sec:ensemble_discussion}

The goal of ensemble modelling needs to be more carefully specified, before a method of comparing ensemble members can be analysed. For example, an error metric would work to compare variation from a ``truth" (e.g. observation) or from a baseline forecast (i.e. before one creates ensemble members by varying parameterisations). We make several observations on ensemble modelling for future use.

A simple use of ensemble modelling would be to sample unknown quantities. Our results suggest that this would need to be done carefully to avoid bias; for example, sampling a range of $\mu$ values would err on the of faster diffusion and earlier monotonicity, because $t_m$ does not change linearly with $\mu$. Furthermore, one should be wary of determining an ``average" from an ensemble;  simply averaging PSD values at each $L$ across many ensemble members would not be meaningful when one needs to consider the whole domain first. This domain dependence raises a further problem;  given an $L$-dependent $D_{LL}$, there is no clear way to compare simulations with a different outer boundary location or condition.

Finally, we note that one possible goal for ensemble modelling could be to characterise the influence of chaotic or stochastic processes. This would need to be carefully thought through, using the inherent properties of such a complex system; simply sampling from the initial conditions above (or from similar values of $D_{LL}$, \cite{Thompson2020}) is not likely to represent either a chaotic or stochastic underlying nature, but only to reinforce any bias towards higher amounts of diffusion. Poorly defined averaging of underlying properties may be why existing diffusion coefficients vary drastically \cite{Murphy2023}. 

In \cref{sec:metricchoice} we motivated $t_m$ by considering the properties required of a metric to analyse our ensemble. We note that our additional, analytic tools $\mathcal{E,N}$ meet the proposed initial requirements (i.e. excluding the requirement for insensitivity to the total particle population) and suggest that these requirements may be useful in finding other qualitative tools for analysing ensembles [A1].

We conclude that $t_m$ and $\mathcal{N,E}$ are good tools for qualitatively understanding what an ensemble is doing, but not necessarily a good tool for comparing ensemble members for modelling or forecasting purposes [P1a;S3b].

\subsection{The Simulation Outer Boundary}\label{sec:Louter_discussion}

We used both Neumann and Dirichlet outer boundary conditions as both have physical motivations. \old{We tested multiple $L_{outer}$ options as the true radiation belt outer boundary is both poorly defined, since particles don't stop existing beyond the last closed drift shell and poorly represented in current models, which often place the outer boundary where observations exist - far short of the true outer boundary. }\new{We tested multiple $L_{outer}$ options as the true radiation belt outer boundary is both (a) poorly defined and (b) poorly represented in current models. The radiation belt outer boundary is poorly defined because particles don't stop existing beyond the last closed drift shell, making it difficult to identify from observations, and because the last closed orbits themselves are not clearly defined; they may be `split' by drift-orbit bifurcations \cite{Desai2021b,Huang_J2022}. The outer boundary can be poorly represented in current models for different reasons, including that this outer boundary changes in time, and that for practical reasons the outer boundary is often placed where observations exist in order to drive that outer boundary. Operational geostationary satellites such as GOES have good coverage around the Earth and for many years and so are very practical for outer boundary conditions, e.g. \cite{Glauert2018,Lee2024}. See \cite{Lee2024} and references therein for other examples of models using in-situ spacecraft to drive the outer boundary. Unfortunately, GOES is situated around $L\sim 6$, far short of the true outer boundary which can vary considerably but is statistically placed at $L\sim8$. These constraints are considered in more detail later in this section, with respect to our results.\ }We have found that both the outer boundary location and condition affects the evolution of the system and the final PSD distribution after a week. The outer boundary condition used in our experiments to identify this effect is very idealised; here we discuss what impacts this may have on outer boundaries used in practice.

$t_m$ showed a clear difference between Neumann and Dirichlet conditions; Neumann conditions reached monotonicity first. $t_m$ also varied with the choice of $L_{outer}$, especially for a Neumann boundary without loss from pitch angle scattering but also for a Dirichlet condition with loss, when the plasmapause is nearby. Using $\mathcal{N,E}$ we also found differences with both outer boundary location and condition. We expected a different long term solution for Neumann and Dirichlet conditions, and found that evolution towards these steady states was very different (i.e. Neumann could reach a lower-$\mathcal{E}$ state, but Dirichlet lost $\mathcal{E}$ more quickly, heading towards their steady state more rapidly). While the initial $\mathcal{E}$ did not change significantly with $L_{outer}$, the evolution of $\mathcal{E}$ did vary; a smaller $L_{outer}$ (and hence a shorter domain) had $\mathcal{E}$ that diverged more with time between Neumann and Dirichlet outer boundary conditions. Loss mitigated, but did not remove, the differences between simulations with outer boundary location and condition.  We conclude that if these options give different dynamics and different PSD distributions over the week, then we need to find the correct boundary conditions. [P1c,S1a,S1b]

Typically, models of the outer radiation belt use a Dirichlet outer boundary to make use of spacecraft observations. These are at positions well short of the true outer edge of the radiation belt, for example they may be curtailed to $L_{outer}=5.5$ \cite{Glauert2014} or extrapolated to higher $L$ values \cite{SubbotinShprits2009,Shprits2006b}.\new{\footnote{This is also close to the typical plasmapause location, which interacts with the outer boundary location and condition.}} Unfortunately, most missions are limited to a few years; for a consistent set of flux observations, one must use geostationary data, for example GOES, observed daily between $L=5.9$ and $6.4$ but mapped to a constant $L$ to enable modelling \cite{Glauert2018}. Whilst at first this appears to be more physical than the idealised boundary conditions tested in this work, we have identified that just the inclusion of observations does not remove all the problems associated with the outer boundary [S1d]. 

Either outer boundary condition, imposed incorrectly, can correspond to erroneous sinks or sources. For example, in our experiments, a Dirichlet condition where the outer boundary value is higher than in the bulk of the simulation domain represents an infinite source of material. Since our Neumann boundary experiments have a plateau that can freely rise, this actually represents an increased source at the outer boundary - which is also not physical. Using a data-driven outer boundary (such as used in most real-life models) would reduce the imposition of unphysical sinks and sources, but we have shown that it does not resolve the problem as these boundaries are placed at a low $L_{outer}$ that curtails the domain, resulting in different evolution (as we are removing the stronger high-$L$ diffusion). The effect of this curtailment changes with outer boundary condition (see e.g. the diverging $\mathcal{E}_t$ in \cref{fig:Louter_NE}(c,e)). From \cref{fig:Louter_par_loss} it's clear that once one includes loss from pitch angle scattering, the evolution differences between outer boundary conditions is reduced, but the shape of the distribution still changes with $L_{outer}$, as $t_m$ is not independent of $L_{outer}$. Using an outer boundary location determined by the different trajectories of different spacecraft is not ideal - the model corresponding to each spacecraft would then have a different $L_{outer}$, and each of these models would respond differently, reaching monotonicity at different times, because outward radial diffusion is not being properly captured.[P1c,S1b]

Would using observations at every timestep sufficiently constrain the simulation so as to remove the variability we see when changing $L_{outer}$ and the outer boundary condition? This is not clear.  It may be that using observations approximates the missing outer boundary processes well enough\new{; results from \cite{Lee2024} using both Dirichlet and Neumann outer boundaries at several outer boundary locations show that using GOES data is clearly superior to a Neumann outer boundary to capture long-term behaviour.}\new{\ 
Our results suggest that the poor performance of the Neumann boundary in \cite{Lee2024} represent inadequate characterisations of the constraining physics, and subsequently unphysical PSD behaviour (i.e. a rising plateau)}
. However, whilst relying on observations to correct improper boundary conditions may perform well for event reproduction (i.e. hindcasting), observations are naturally not available for the future. This may limit how far in advance we can model radiation belt behaviour. Neither Neumann nor Dirichlet boundary conditions are suitable for predictive purposes. Ideally we would not have to pick between these when we do not have clear values with which to constrain this outer boundary (as is the case here).  Currently, we are reduced to comparing empirically which simulation settings account for variation across more orders of magnitude, rather than solely using physical motivations\new{\ to understand the resulting uncertainty} [S1d]. 
%
%
%
%

Options would therefore be to used mixed (Robin) boundaries that relate the flux from the domain to the outer boundary values, and/or some way to include the observations in the middle of the domain such as data assimilation or source terms. Methods such as these are already under investigation to resolve the fact that current modelling misses processes such as dropouts \cite{Daae2011, Cervantes2020} [S1a,S1c]. 

However, even a Robin boundary condition would not fix the fact that a true outer boundary location is both difficult to define and difficult to find. One could find the empirical extent of the highly-charged particles. However, this is not the same as the last closed drift shell (LCDS), which is the outer limit of adiabatically trapped particles. 
Particles within a closed drift shell will continue to drift on their path (of constant magnetic field) around the Earth. However, if a drift shell is open then sections of the drift path lie on open field lines outside the magnetosphere. Here, particles can be lost to the solar wind. The last closed drift shell is the last point in which particles are trapped rather than being lost to the solar wind. Therefore the LCDS could be considered the ``true" edge of the diffusion domain (but not the edge of the particle population) [S1a].

There is significant uncertainty in the location of the LCDS as modelled using state-of-the-art global magnetosphere models (e.g. \cite{Albert2018}) and there is significant variability in the location of the LCDS (e.g. \cite{Bloch2021}) due to motion of the magnetopause and spatiotemporal variability of the magnetic field in Earth's outer magnetosphere. \cite{Bloch2021} estimated the typical outer boundary to be at $L^* \sim 8 R_E$ - significantly more distant than models used in practice [S1d]. 

In this work we have shown that in simulations, both the outer boundary condition and location change the rate of evolution and the shape of the phase space density distribution. We have argued that using a curtailed domain to set a Dirichlet outer boundary may remove these problems for historical event studies, but are not suitable for predictive purposes. Finally, as radial diffusion is about the diffusion of particles across different drift paths, radial diffusion is not well defined when drift paths are open. Therefore a theoretical limit to the radiation belts is the last closed drift shell (LCDS). However, this is (a) a dynamic boundary, (b) difficult to identify in practice, (c) not a closed outer boundary (i.e. particles can be lost to or gained through it) and (d) still an approximation, as in reality the last closed drift orbits are not uniquely defined. It may be impossible to set a ``true'' outer boundary; in the meantime we do not know what level of accuracy is needed in the outer boundary location and conditions to adequately reflect the radiation belts.
We conclude that significant work is required to identify reasonable outer boundary conditions and location for modelling; the outer boundary in real life is very variable, and we have shown that simulations are sensitive to several outer boundary choices [S1a,c,d].



\begin{notestocoauthors}
    {\color{purple}here is the diagram from Clare and my converstaio, if that helps}
\begin{figure}
    \centering
    \includegraphics[width=0.6\textwidth,angle=-90]{figs/diagram7.jpg}
    \caption{Diagram to emphasise difference between using LCDS vs number of particles for the outer boundary. Where trapped vs untrapped particles end. Ideally also PSD values at each of these locations, for trapped and untrapped. Add: injections at high L from substorms.}
    \label{fig:diagram7}
\end{figure}

\end{notestocoauthors}

\subsection{Improving $D_{LL}$ vs improving PSD  distribution}\label{sec:gradients_discussion}
\begin{notestocoauthors}
    {\color{red}\textbf{Clare:} If you have any good comments, or references here for anyone trying to find PSD, please add!}
\end{notestocoauthors}

The overwhelming recent focus to improve radial diffusion is through the diffusion coefficient $D_{LL}$, for example the theory behind $D_{LL}$ \cite{LejosneAlbert2023}, the strength of the electromagnetic perturbations driving $D_{LL}$ (both generally, \cite{Bentley2020} , or for event specific diffusion coefficients \cite{Silva2022,Li2020}  ), parameterisations of $D_{LL}$ for use operationally \cite{Murphy2023}, the effect of plumes on diffusion coefficients \cite{Sandhu2023},   or even radial diffusion vs radial transport \cite{Osmane2023}. However, our results suggest that the gradients of the underlying phase space density distribution may have more effect on the radial diffusion \new{. This is an unexpected result to radiation belt modellers and will need to be tested further, for example using more complex diffusion coefficients and/or expanding from just radial diffusion to include other radiation belt dynamics }[A2].

Comparatively less effort has gone into finding the underlying PSD distribution. Typically, this is built up on a case-by-case basis for event studies, or to specifically understand the mechanism behind enhancements, rather than as a foundation for radiation belt modelling. There are many difficulties with the construction of radial PSD profiles, particularly as these are made from observations which are sparse in time and space, often on scales much slower than enhancements \cite{Kim2023b}. \cite{Turner2012} includes a statistical study and emphasises the need to include error and uncertainty. Our results above (\cref{sec:gradients}) indicate that it is the gradients in the radial profile, rather than the exact $D_{LL}$, that determines radial diffusion, to several orders of magnitude. We have not explored the extent of this result: if there is a limit to this with a stronger L dependence, a longer domain or the choice of boundary condition. Nevertheless, as radial diffusion is the large scale, bulk mechanism behind radiation belt evolution, and recent PSD profiles are shown to regularly contain gradients from enhancements \cite{Chen2020,Kim2023a,Kim2023b}, finding the radial PSD profiles may be a more valuable future route of study; the form of $D_{LL}$ may be less important than getting the gradients of local enhancements right.\new{\ If this is the case, then }\old{
However, }given that the uncertainty in $D_{LL}$ is also of orders of magnitude\new{\ (see discussion in \cref{sec:Intro,sec:Background})}, improved $D_{LL}$s are still likely to result in improved radiation belt modelling. \new{Either way, this work demonstrates that analytical tools based on the principles here could result in valuable ways to test radiation belt models and to quantify the impact of model components such as gradients and diffusion coefficients [P1a; A1]}




\new{\subsection{The Role of the Plasmapause in Loss and Radial Diffusion}}

\new{The plasmapause is already known to contribute to radiation belt dynamics, especially through wave-particle interactions faster than radial diffusion (i.e. affecting the first and second adiabatic invariants). Plasmaspheric structure is not typically considered a significant component to radial diffusion outside the loss due to pitch-angle scattering, which was the original expectation here. However, we have shown that although the loss from pitch-angle diffusion dominates over radial diffusion, the spatial \emph{limit} (in $L$) of this loss has significant implications for the PSD evolution.}

\new{Furthermore, plasmaspheric structure may can also affect the radial diffusion directly in a way that is not incorporated to radiation belt models today. Recent work has shown that ULF waves can vary in structure when plasmaspheric plumes arise, which will consequently affect on the radial diffusion \cite{Sandhu2021c,Elsdon2022b,Sandhu2023}. Our work has demonstrated that even a simple plasmapause interacts with the initial morphology of the PSD distribution (especially the location of the central peak of an enhancement) and the outer boundary of the simulation to produce very different final PSD profiles with different maxima. A more realistic scenario would include (a) a plasmapause varies azimuthally around the Earth, (b) plume structure instead of a single plasmaplause and (c) increased radial diffusion inside that plume. Developing the interactions we have identified here for this more realistic scenario may have significant implications for the evolution of electron PSD in the radiation belts in practice.}

\subsection{Limitations of our numerical experiments}
The strengths and weaknesses of our study arise from the same principle: idealised experiments. By examining the fundamentals of radial diffusion modelling we aimed to understand the results of ensembles. To do so, we made many simplifications, which we shall review.

Our experiments showed that the ``background'' initial condition used above should be skewed further to lower $L$, for example the final monotonic distributions shown in \cref{fig:diagram3}, rather than the step-and-bump of our initial conditions. The ``background'' higher PSD at the outer boundary is assumed to be from substorms. Given that the inter-substorm time is relatively fast, with a mode of 3 hours (\cite{FreemanMorley2004,Keiling2022}), this ``background'' level is a reasonable initial condition. We also used a Gaussian to represent an enhancement. However, if the enhancements occur on the order of days, single events (and therefore ``time for an enhancement to diffuse away'') should be replaced by compounded events \cite{Kim2023a}.

We did not explore the result of variance of the inner boundary. While the outer boundary condition and location have been thoroughly explored, they are unrealistic. They do not include observations as per most operational models and despite being able to physically motivate these settings initially, they essentially relate to an unphysical source or sink at the boundary. This poses several questions about how radiation belt modelling should be attempted in future, which are covered in \cref{sec:Louter_discussion}; a different methodology than ours would be needed to investigate these questions, and to examine the relative effect of different driving conditions at the outer boundary, which was out of the scope of this investigation.

The radial diffusion equation \cref{eq:diffusion+loss} is well-known and widely used. Our results confirm that it may not be meaningful to consider radial diffusion without loss from pitch angle scattering. Our equation for loss is very simple, in order to match the number of parameters used in our experiments; more sophisticated parameterisations exist (e.g. \cite{Orlova2016}). To keep our problem tractable, we also only considered a single value for the first and second adiabatic invariants. A downside of our simple loss equation is the lack of time dependence; unlike real life, the amount of particles lost does not vary, except when we set a different plasmapause location. 
Of course, exploring all these options is a difficult problem; in order for the problem to remain tractable, we chose not to vary these. \new{We have also only included loss from the interior of the domain (precipitation). Loss across the magnetopause, also known as magnetopause shadowing, is necessary for realistic radiation belt models but again requires estimates of the radiation belt outer boundary.}

A major choice in this investigation was the Ozeke and idealised versions of $D_{LL}$, \cref{eq:OzekeDLL} and \cref{eq:LnDLL}. Unfortunately we could not use a single $D_{LL}$ throughout the investigation; whilst the Ozeke model is a simple parameterisation that is widely used operationally, it does not admit easy solutions for $\mathcal{N}_t$ and $\mathcal{E}_t$. The parameterisation of Kp is very rough and is a coarse average over many different processes that all affect radial diffusion (e.g. ULF waves, compressions, plumes). Nevertheless, both models are comparable to diffusion coefficients used elsewhere\new{, given the dependence on $L^{6},L^8$ etc multiplied by a proxy for the effect of electromagnetic perturbations, e.g.} \cite{Brautigam2005,Fei2006,LejosneKollman2020}. The assumption that some form of radial diffusion is always ongoing is a reasonable one as there is always some level of ULF waves; \new{a summary of several\ }other limitations of \new{currently applied\ }radial diffusion modelling can be found in \cite{Bentley2019}.\new{\ In this manuscript we have only discussed limitations of the quasilinear techniques typically used, yet recent work indicates that current diffusive models cannot contain all radial mechanisms and nonlinear contributions to radial diffusion are needed\ }\cite{Osmane2023}. 

Despite the large number of idealisations required for this work, we nevertheless reached some general and significant conclusions that will inform future ensemble modelling and suggest that new directions are needed for radiation belt modelling. These are summarised below.




\section{Conclusion}
We used numerical experiments to investigate how initial conditions and basic radial diffusion modelling techniques affect the final phase space density of radiation belt ensembles. As yet, there are few standard methods of analysing ensembles, hence we explore some in this work. Despite the radial diffusion equation arising from the simple heat equation, the space and time dependence of the diffusion coefficient $D_{LL}$ make analysis of this system rather complicated, even while space weather modelling demands are increasing. Our initial goals included defining a radial diffusion timescale and the effect of model settings across ensemble members, yet as part of this work we have identified significant questions in current practice for radiation belt modelling. To aid navigation of this paper, each goal has been explicitly stated in \cref{sec:goals} and tagged throughout.

The key findings are summarised here and briefly expanded below:
\begin{enumerate}
    \item Evolution of the system depended on the outer boundary condition and location. A shorter domain evolved at a different rate than a longer one; this will be due to the $L$ dependence of $D_{LL}$. It is not clear what outer boundaries should be used and this may have consequences for modelling the radiation belts [P1c;S1].
    \item Using an analytical quantity (the energy moment, or norm $\mathcal{E}$) we found that the gradient of the phase space density distribution contributed more to the evolution of the system than the diffusion coefficient $D_{LL}$ [A2].
    \item Flaws in typical metrics included (a) threshold-based metrics which gave results highly threshold-dependent, and (b) the requirement of a $\frac{1}{L^2}$ Jacobian factor due to the co-ordinate system [P1a].
    \item Time to monotonicity $t_m$ and mass / energy moments $\mathcal{N},\mathcal{E}$ were developed to analyse radial diffusion models [P1a].
    \begin{itemize}
        \item These metrics were appropriate because they consider the whole domain and are $L$ dependent,
        \item $t_m$ is intuitively interpretable as time for an enhancement to diffuse away [S2],
        \item the system can be considered as continually moving to a low energy moment (or $L_2$ norm) $\mathcal{E}$ state [A1],
        \item $\mathcal{N},\mathcal{E}$ could be easily adapted to other radiation belt models [S1].
    \end{itemize}
    \item Loss from pitch angle scattering generally dominated over radial diffusion [S4].
    \item \new{Taking an average over ensembles where the enhancement location $\mu$ varied in $L$ would result in a PSD biased towards additional radial diffusion; linear increases in $\mu$ result in nonlinear decreases in $t_m$. This analysis could be extended to find how mass and energy density $\mathcal{N},\mathcal{E}$ vary across gradually changing ensemble members}\old{Ensembles varying the enhancement location would be biased towards additional radial diffusion} [P1b,S3].
\end{enumerate}

The methodology of this study was to perturb simulations, selecting ensemble members based on sampling idealised conditions from Earth's radiation belts. Initial conditions were a ``background'' phase space density (PSD) distribution, plus a localised enhancement. These properties were varied and their impact on the evolving PSD distribution assessed using time to monotonicity $t_m$ and two analytic mass- and energy-like quantities $\mathcal{N,E}$. Multiple physical outer boundary conditions and locations were tested, using both empirically fitted and idealised radial diffusion coefficients. Finally, loss from pitch angle scattering was included. Throughout this paper, the impact of all these factors on the PSD have been extracted. Here we summarise significant findings and important discussions.

The final PSD, and evolution towards that state, varied with both the outer boundary condition and location. Both constant flux (Neumann) and fixed value (Dirichlet) outer boundaries can be physically motivated, although neither can be easily implemented to reflect real-life conditions. Current operational methods generally involve either a diffusion domain shortened to where observations are available, or extrapolations from this point to a distant outer boundary. We have shown that while observations could constrain simulation errors from a curtailed boundary for historical events, this is far from assured when modelling future behaviour. The interplay of domain-dependent $D_{LL}$ and outer boundary condition and location as an additional source of error has not been heretofore considered. We suggest several ways forward, including mixed boundaries and data assimilation. Our work also suggests that identifying an outer edge to the real radiation belts - currently poorly defined and difficult to identify from observations - will have significant implications for modelling. In particular, as modelling of the radiation belts becomes more realistic, the outer boundary location (and therefore the size of the domain) will vary more in time in our model [P1c;S1a;S1b;S1c;S1d;S2e].



The bulk of previous work on improving radial diffusion estimates has focused on finding and characterising $D_{LL}$. By comparing the components of $\mathcal{E}_t$, which determines the ongoing evolution of the system, we find that it is instead the gradients of the PSD distribution that determine the ongoing amount of diffusion.\new{\ This result was for an idealised 1-d (radial only) diffusion and idealised diffusion coefficientis, and needs to be examined in more realistic contexts.} This result also emphasises the need to integrate any comparative measures across the entire domain to capture the full impact of the $L$ dependence of radial diffusion. Diffusion is not well characterised by either the largest or smallest $D_{LL}$ in the domain [A2;A3].

Upon analysis of our measures $t_m$ and $\mathcal{N,E}$, we find that their effectiveness is partly due to their relationship to gradients and hence the morphology of the PSD distribution. Our two measures are related, yet complementary. Together they include information on both the system state and the ongoing diffusion. They are both interpretable; $t_m$ has a clear physical interpretation (how long until an enhancement has been completely diffused away) while the components of $\mathcal{N}$ and $\mathcal{E}$ describes the mass in the simulation and the ongoing evolution to a steady state respectively. Using an idealised diffusion coefficient in $\mathcal{N,E}$ enables an explicit comparison of the inner boundary, outer boundary and loss terms on the system.  A limitation of $\mathcal{N,E}$ is the requirement of an idealised $D_{LL}$, whilst ideally $t_m$ should be adapted to enable better application for loss from pitch angle scattering. These are excellent tools for qualitatively understanding what an ensemble is doing, but they are not error metrics. For error metrics it is likely that $L$ will need to be accounted for, e.g. with the Jacobian $1/L^2$. Both measures are domain- and boundary type-dependent; this is a desirable property as it highlights the fact that for radial diffusion, different domain sizes and boundary types are effectively modelling different physical systems [P1a;P1b;P2a;S2;S3;A1].

Using $t_m$ and $\mathcal{N,E}$ all the components of a typical radial diffusion model were compared. Loss from pitch-angle scattering (using an extremely simple loss model) generally had more effect on the PSD distribution than diffusion, although this was highly dependent on the location of the plasmapause. A more distant plasmapause results in so much more loss that the dynamics could change totally, and the system never be able to reach a monotonic state. Of the initial condition, step size parameterised by $B$ (corresponding to the quiet background PSD distribution) and enhancement location $\mu$ had the greatest effect on the timescale of diffusion and the state of the system; these results agree with our findings that gradients control the diffusion. Roughly, we found that $
    \text{loss} > \text{outer boundary} > \text{initial condition} > \text{inner boundary}
$ for the impact on evolution, by order of magnitude; extreme values change this ordering. The initial condition controlled loss at the inner boundary [P2c;S3;S4;A1].

To summarise, in this paper we have presented a methodology to analyse the components of a numerical model of radial diffusion in Earth's radiation belt and compare the impact of those components on the shape and evolution of the particle phase space density distribution. Our work emphasises the need to consider the entire modelling domain when comparing or analysing radial diffusion simulations. Furthermore, we find that the $L$-dependence of the diffusion coefficient results in a model that is domain dependent; curtailing the outer boundary or using inappropriate boundary conditions effectively models very different systems. We make several suggestions for the outer boundary in future models. We find that it is the gradients of the phase space density that mainly control diffusion, contrary to the focus of most radial diffusion studies on $D_{LL}$. 

\new{
\section{Supplementary Materials}
The supplementary materials contain plots for all numerical experiments run for this paper, i.e. the time to monotonicity ensembles for each parameter $A,B,\mu,\sigma, L_{outer}, L_p$ with and without loss, and the calculated quantities $\mathcal{E},\mathcal{E}_t,\mathcal{N},\mathcal{N}_t$ for each parameter also. In the main body of the manuscript, selected results were shown to convey significant results.
}

\begin{notestocoauthors}
{\color{purple}

\section{Conclusions - bulleted}
I find these sooo much easier than text. I used them to write conclusion. Should I do something with them - e.g.  supplementary materials? blog? or just cut?

\subsection*{Outer boundary location and condition}
  \begin{enumerate}
    \item We should be using an outer boundary location and condition that reflects the radiation belts.
    \begin{enumerate}
        \item Both our measures are dependent on outer boundary location and condition.
        \begin{enumerate}
            \item a smaller $L_{outer}$ (and hence a shorter domain) had $\mathcal{E}$ that diverged more with time with Neumann than Dirichlet outer boundary condition. 
        \item $t_m$ varies significantly with $L_{outer}$ for Neumann without loss, and slightly with both Neumann and Dirichlet OBC when there is loss
        \end{enumerate}
        \item The choice of outer boundary location and condition change the shape of the PSD distribution, both with and without loss.
        \item Loss may be more significant than outward radial diffusion and may mitigate, but not remove, the problem of unphysical boundary conditions.
        \begin{enumerate}
            \item With loss, there can be additional differences if the plasmapause location is too close to the outer boundary.
            \item $t_m, \mathcal{E}$ still have differences with outer boundary location and condition, even if loss dominates evolution (i.e. if loss dominates $\mathcal{E}_t$)
        \end{enumerate}
        \item Both conditions can be physically motivated, but neither of our implementations are really physical
        \begin{enumerate}
            \item They correspond to implausible sinks/ sources
            \item With Neumann and Dirichlet outer boundary conditions, we would expect different steady states
        \end{enumerate}
        \item Current implementations (a Dirchlet outer boundary, set by observations) seem more physical but since they curtail the domain to an artificially short $L_{outer}$, some of the important diffusion (i.e. high $L$ and hence high $D_{LL}$) is omitted, resulting in different $\mathcal{E}_t$ and hence different evolution of the PSD
        \item We suggest mixed boundary conditions and/or data assimilation for the internal part of the domain.
    \end{enumerate} 
           \item The outer edge of the radiation belt itself is poorly defined \textbf{and} difficult to identify from observations. We already knew this, but the domain dependence of $D_{LL}$ means that finding the anser to this will materially affect our modelling.
         %
\end{enumerate}

\subsection*{Gradients}
    \begin{enumerate}\setcounter{enumi}{2}
    \item Gradients in the PSD are really important, quite likely more important than $D_{LL}$
    \begin{enumerate}
        \item $\mathcal{E} $ and $t_m$ tell us very different things about the gradients; the effect of gradients on the ongoing evolution of the system, and whether a localised enhancement has been diffused into the background particles.
        \item the form of $D_{LL}$ may be less important than getting the gradients of local enhancements right.
        \item Diffusion is not limited by the smallest diffusion coefficient, but by the gradients across the entire domain; large enough $L_{outer}$ may mean the diffusion coefficient starts to dominate.
    \end{enumerate}
\end{enumerate}

\subsection*{Our two measures}
\begin{enumerate}\setcounter{enumi}{3}
    \item We used two complementary techniques to compare radial diffusion ensembles
    \begin{enumerate}
    \item \textbf{Time to monotonicity $t_m$} 
    \begin{enumerate}
        \item $t_m$ represents our physical expectations; our intuition that after a localised enhancement the PSD will eventually relax to monotonic distribution. 
        \item It’s pretty robust and therefore great to use across an ensemble to examine our model methods vs the physics. It fails when monotonicity will never be reached.
        \item It can be applied to any radial diffusion equation, but needs to be adapted when including loss (perhaps using convexity arguments)
        \item $t_m$ is domain dependent and boundary type dependent. This makes sense as our experiments turn out to be dependent on these too.
    \end{enumerate}
    \item $\mathcal{N}$ tells us about the mass in the simulations. Terms $\mathcal{N}_t$ tell us about the dominant factors for the changing mass.
    \item $\mathcal{E}$ is the $L_2$ norm; an energy-like quantity that shows us the ongoing evolution to a steady state. Terms $\mathcal{E}_t$ show us the dominant factors in this evolution.
    \item $\mathcal{N,E}$ need a simple description of diffusion coefficient, we used $D_{LL}=D_0 L^n$
    \item Error metrics for measuring ensemble reproductions of real-life observations has not been resolved, but we discuss potential avenues in \cref{sec:ensemble_discussion}
    \item Initially, we searched for domain \emph{in}-dependent measures; but when the diffusion coefficient $D_{LL}$ is domain dependent, so should our method of analysis.
    \end{enumerate}
\end{enumerate}

\subsection*{Loss and plasmapause}
    \begin{enumerate}
  \setcounter{enumi}{4}
    \item Loss is important, location of plasmapause is importnat
    \begin{enumerate}
        \item Plasmapause location relative to enhancement and to outer boundary of domain is important 
        \item A more distant plasmapause has so much more loss that this can totally change the dynamics
        \item A plasmapause at higher L than an enhancement could mean that the experiment does not reach monotonicity
    \end{enumerate}
\end{enumerate}

    \subsection*{Misc other to sort}
    \begin{enumerate}
  \setcounter{enumi}{5}
    \item you have to consider the whole domain.
    \item Summarise main contributing initial condition parameters, and say that boundary conditions are more important
    \item initial condition controls IB more than OBC. 
\end{enumerate}

}
\end{notestocoauthors}

\begin{acknowledgments}
We would like to thank the reviewer for their thoughtful comments and attention to detail in making this manuscript more accessible.
\begin{notestocoauthors}
    
{\color{purple}
CRediT system (https://credit.niso.org)
\begin{itemize}
    \item Conceptualization: SNB, RT, CEJW
    \item Data curation: SNB (RT?)
    \item Formal analysis: SNB
    \item Funding acquisition: SNB
    \item Investigation: JS, SNB
    \item Methodology: SNB, RT, JS, DJR, CEJW
    \item Project administration: SNB
    \item Resources: RT
    \item Software: RT, JS, SNB
    \item Supervision: SNB, RT
    \item Validation: SNB, if she uploads code etc
    \item Visualization: SNB, JS
    \item Writing – original draft: SNB 
    \item Writing – review \& editing: DJR, RT, CEJW
\end{itemize}

We found that WHAT FOR DIRICHLET AS LENGTH CHANGES – WHAT DIFFUSES MORE IN MASS/ENERGYS

L=7.5 lost 46.181 (0.766

Dirichlet: longer domain -> Starts with more mass, does it lose more or less? Shorter domain is reaching “most diffused state” more quickly

}
\end{notestocoauthors}

JS was supported by a RAS Summer Bursary. RLT was supported by the Engineering and Physical Sciences Research Council (EPSRC) (Grant no. EP/L016613/1). SNB was supported in part by STFC Grant ST/R000921/1. CEJW was supported in part by STFC ST/W000369/1 and NERC NE/V002759/1. Both SNB and CEJW were supported in part by STFC ST/X001008/1.

\end{acknowledgments}

\section*{Data Availability Statement}
\old{Data generated in support of this paper will be uploaded to an online repository with a DOI after first review.}\new{The data that support the findings of this study are openly available in Zenodo at \url{https://zenodo.org/doi/10.5281/zenodo.13314166} using the Python notebook on Zenodo/GitHub \url{https://zenodo.org/doi/10.5281/zenodo.13336681} \cite{Bentley-dataset-2024,rad-diff-POP}.}



\nocite{*}

\bibliography{ICpaper}

\end{document}